\documentclass[journal=accacs,manuscript=article,layout=twocolumn]{achemso}
\usepackage{chemformula} 
\usepackage[T1]{fontenc} 
\usepackage{xr} 
\usepackage{gensymb} 
\usepackage{caption}
\usepackage{subcaption}
\usepackage{tabularx}

\usepackage{hyperref}

\makeatletter
\newcommand*{\addFileDependency}[1]{
\typeout{(#1)}
\@addtofilelist{#1}
\IfFileExists{#1}{}{\typeout{No file #1.}}
}\makeatother

\newcommand*{\myexternaldocument}[1]{%
\externaldocument{#1}%
\addFileDependency{#1.tex}%
\addFileDependency{#1.aux}%
}

\myexternaldocument{LTO_Positron_SI}


\author{Yute Chan}
\affiliation{Chair for Theoretical Chemistry and Catalysis Research Center, Technical University of Munich, Lichtenbergstr. 4, 85747 Garching, Germany}
\alsoaffiliation{Fritz-Haber-Institut der Max-Planck-Gesellschaft, 14195 Berlin, Germany}
\altaffiliation{These authors contributed equally to this work} 
\author{Cristina Grosu}
\altaffiliation{These authors contributed equally to this work}
\affiliation{Chair for Theoretical Chemistry and Catalysis Research Center, Technical University of Munich, Lichtenbergstr. 4, 85747 Garching, Germany}
\alsoaffiliation{Institute of Energy Technologies (IET-1), Forschungszentrum J\"ulich, 52425 J\"ulich, Germany}

\author{Matthias Kick } 
\affiliation{Chair for Theoretical Chemistry and Catalysis Research Center, Technical University of Munich, Lichtenbergstr. 4, 85747 Garching, Germany}
\alsoaffiliation{Fritz-Haber-Institut der Max-Planck-Gesellschaft, 14195 Berlin, Germany}

\author{Peter Jakes}
\affiliation{Institute of Energy and Technologies (IET-1), Forschungszentrum J\"ulich, 52425 J\"ulich, Germany}
\author{Stefan Seidlmayer}
\affiliation{Heinz Maier-Leibnitz Zentrum (MLZ), Technical University of Munich, Lichtenbergstr. 1, 85748 Garching}
\author{Thomas Gigl}
\affiliation{FRM II at Heinz Maier-Leibnitz Zentrum (MLZ), Technical University of Munich, Lichtenbergstr. 1, 85748 Garching}
\author{Werner Egger}
\affiliation{Institut f\"ur  Angewandte Physik und Messtechnik LRT2, Werner-Heisenberg-Weg 39, D-85577 Neubiberg} 
\author{R\"udiger-A. Eichel}
\affiliation{Institute of Energy and Technologies (IET-1), Forschungszentrum J\"ulich, 52425 J\"ulich, Germany}
\alsoaffiliation{RWTH Aachen University, Institute of Physical Chemistry, D-52074 Aachen, Germany}
\author{Josef Granwehr} 
\affiliation{Institute of Energy and Technologies (IET-1), Forschungszentrum J\"ulich, 52425 J\"ulich, Germany}
\alsoaffiliation{RWTH Aachen University, Institute of Technical and Macromolecular Chemistry, D‐52074 Aachen, Germany}
\author{Christoph Hugenschmidt}
\affiliation{FRM II at Heinz Maier-Leibnitz Zentrum (MLZ), Technical University of Munich, Lichtenbergstr. 1, 85748 Garching} 
\author{Christoph Scheurer} 
\email{scheurer@fhi.mpg.de}
\affiliation{Chair for Theoretical Chemistry and Catalysis Research Center, Technical University of Munich, Lichtenbergstr. 4, 85747 Garching, Germany} 
\alsoaffiliation{Fritz-Haber-Institut der Max-Planck-Gesellschaft, 14195 Berlin, Germany}
\date{\today}
\title
  {The Origin of Enhanced Conductivity and Structure Change in Defective $Li_{4}Ti_{5}O_{12}$ or Blue-LTO : a study combined theoretical and experimental perspectives}

\abbreviations{PALS, CDBS, DFT+U, LTO, Polaron}

\begin{document}

\begin{tocentry}
\begin{center}
    \includegraphics[width=9cm,height=3.5cm]{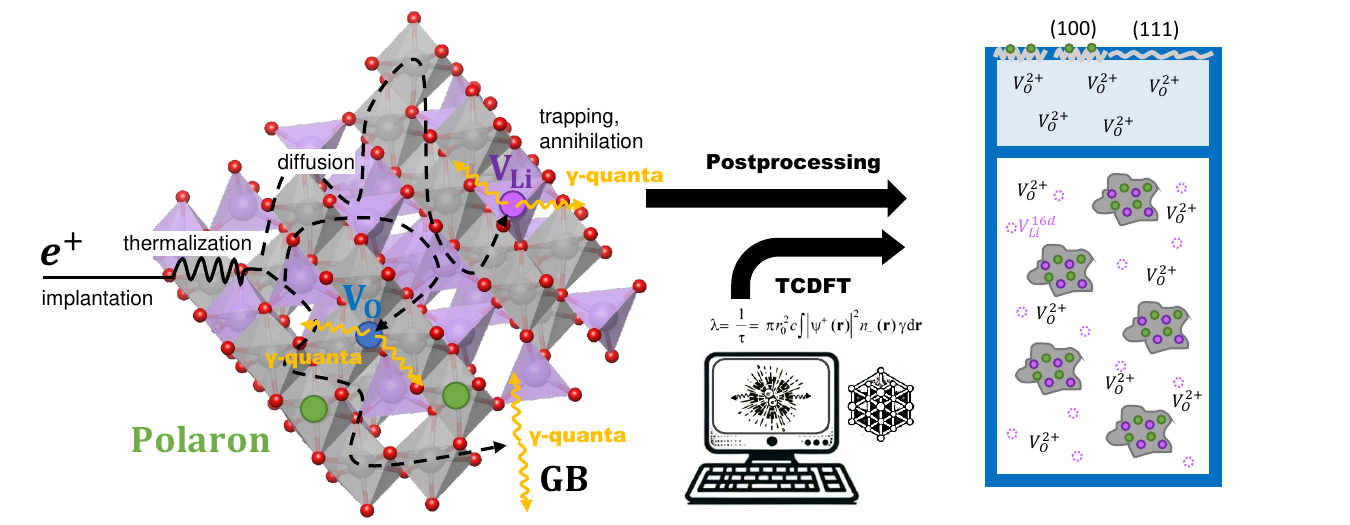}
\end{center}




\end{tocentry}
\begin{abstract}
The spinel \ch{Li4Ti5O12} (LTO) has emerged as a promising anode material for the next generation of all-solid-state Li-ion batteries (ASSB), primarily due to its characteristic "zero strain" charge/discharge behavior and exceptional cycling stability, which significantly prolongs battery lifespan.
Pristine LTO, however, is hindered by poor ionic and electronic conductivity. By employing tailored sintering protocols that create oxygen vacancies, a high-performing, blue LTO material is achieved. It has been proposed that the increased electronic conductivity could stem from vacancy-induced polarons. Yet, detailed insights into polaron stability, distribution, and dynamics within both the LTO bulk and surface have remained elusive due to limited information on structural changes.
Utilizing Positron Annihilation Lifetime Spectroscopy (PALS) and Coincidence Doppler Broadening Spectroscopy (CDBS), in conjunction with Two Component Density Functional Theory (TCDFT) with the on-site Hubbard U correction, enables us to probe the depth profile of defect species introduced by sintering in a reductive environment. Our research provides direct evidence of oxygen vacancy formation within the subsurface region, an inference drawn from the observation of \ch{Ti^{3+}}. Our investigation into \ch{Li^{16d}} vacancy formation within the bulk region uncovers the interactions between mobile species, namely Li-ions and polarons. Furthermore, we delve into the polaron stability on the LTO surface, offering an explanation for the superior performance of the (100) facet exposed LTO nanoparticle, as compared to its (111) exposed counterpart.
\end{abstract}

\section{Introduction}
The ever-increasing demand for portable electronic devices, electric vehicles, and renewable energy storage systems has significantly escalated the requirement for advanced battery technologies\cite{Energy_Storage_Review, Beyond_LIBs}. Lithium-ion batteries are widely recognized as a promising energy storage solution owing to their high energy density, extended cycle life, and low self-discharge rate. Despite their inherent advantageous features, there still exists room for enhancing the performance and safety of lithium-ion batteries to meet the demands of potential applications. This can be achieved by addressing various challenges, including dendrite formation, lithium plating, solid electrolyte interphase (SEI) formation, and active materials breakdown/dissolution - all of which contribute to issues such as short-circuiting, poor conductivity, capacity fading, and voltage fading\cite{LIB_dendrite_formation, SEI_LIBs, LIB_aging}. These issues are closely associated with the underlying material composition and atomistic structure, which can be optimized through doping\cite{dendrite_doping}, coating\cite{dendrite_coating}, SEI engineering \cite{SEI_Li_depletion}, defect engineering\cite{ LIB_defect_1, LIB_defect_2}, and facet modification\cite{LIB_facet_1, LIB_facet_2} strategies. Thus, advancements in the fundamental understanding of the atomistic level changes introduced by these processes can lead to the development of superior lithium-ion batteries with improved capacity, efficiency and safety.

In this context, lithium titanate (\ch{Li4Ti5O12}, LTO) has emerged as a promising anode material for lithium-ion batteries due to certain outstanding features, such as excellent thermal stability, negligible volume change during lithium insertion/extraction, and high rate capability\cite{intro_LTO_1, intro_LTO_2, intro_LTO_3}. Despite these advantages, the spinel structure of LTO is inherently burdened with low electronic and ionic conductivity, posing a limitation to its overall electrochemical performance. To overcome these limitations, one approach is creating oxygen vacancies in the pristine LTO through defect engineering, which can enhance both electronic and ionic transport properties\cite{Blue_LTO_Li_diffusivity,blue_lto_oxygen_defecct_1, blue_lto_oxygen_defecct_2, blue_lto_oxygen_defecct_3, Matthias_JPCL, Matthias_polaron_assisted}.Moreover, near-surface defect engineering suffices to considerably affect bulk properties of LTO \cite{Schleker2023}, facilitating a chemical equilibrium between electrolyte and the bulk of LTO\cite{Schleker2023_2}. However, the precise structural changes occurring during the modification process remain to be thoroughly explained. Furthermore, the likelihood of other factors contributing to the observed enhancements cannot be ruled out, as indicated by the varied lithium-ion diffusion rate increases, ranging between five orders of magnitude, as reported in different studies\cite{Blue_LTO_Li_diffusivity, blue_lto_oxygen_defecct_2,Rho2004,Zaghib1999,Bach1999,Kavan2003,Qiu2014,Feckl2012}. Importantly, these studies have consistently identified the presence of Ti(III), often interpreted as indirect evidence for the formation of oxygen vacancies. Comprehensive and direct defect measurements could significantly enhance our understanding of the intricate mechanisms underlying these phenomena.  On the other hand, the noticeable improvements in both electronic conductivity and capacity observed in LTO nanoparticles showcasing both exposed (100) and (111) facets—as compared to those exclusively featuring the (111) facet—suggest a significant influence of surface orientation on the storage and transport of charge carriers\cite{Pal_111vs100}. Despite these observations, the precise underlying mechanism continues to remain elusive.


In this study, we employ a combination of TCDFT+U simulations and experimental techniques such as Positron Annihilation Lifetime Spectroscopy (PALS) and Coincidence Doppler Broadening Spectroscopy (CDBS), to  gain a more profound understanding of the roles played by oxygen vacancies, other defects, and facet effects within the LTO lattice. We discuss their impact on the electronic and ionic transport properties. 
Furthermore, we scrutinize the defect species and their distribution patterns both prior to and following defect engineering. This process involves the application of a reductive atmosphere (\ch{Ar}/\ch{H2}) at high-temperature treatment, traditionally believed to exclusively create oxygen vacancies. 
Our findings significantly advance the fundamental understanding of the intricate interplay between defect structures, mobile species distribution and the electrochemical performance in LTO-based lithium-ion batteries. Moreover, they provide valuable insights for devising advanced strategies to engineer and optimize LTO and other battery materials, paving the way for improved capacity, efficiency, and safety in next-generation energy storage applications. We also demonstrate how the integration of DFT+U simulations with experimental techniques such as PALS and CDBS, presents a promising approach to study defect structures in complex materials. 
\section{Methodology}
\begin{figure}[h]\label{LTO_PAS}
\begin{subfigure}{0.5\textwidth}
\includegraphics[width =\columnwidth]{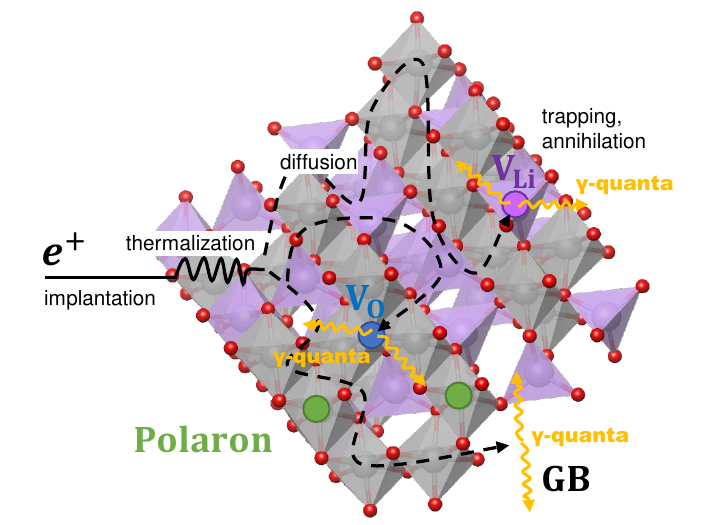} 
\caption{Positron annihilation in Li$_{4}$Ti$_{5}$O$_{12-x}$ (LTO).The figure shows the main processes when the positron interacts with LTO, before its annihilation, in one of the suitable sites, \textit{i.e } vacancy (V), open volume (O-vol) or at the grain boundary (GB), with the release of the two $\gamma$-quanta.  }
\end{subfigure}
\end{figure}

Positron Annihilation Lifetime Spectroscopy (PALS) is a powerful experimental technique capable of providing valuable information on the defect and local structure properties in materials\cite{PALS_polymer, PALS_semiconductor, PALS_TiO2_hydrogenated}. The operational principle of PALS is anchored on detecting annihilation events between injected positrons —electron antiparticles— and the sample's intrinsic electrons. Once introduced into the sample, positrons can be trapped in open-volume defects such as vacancies or voids, resulting in a distinctive lifetime. By analyzing the positron lifetime spectrum, PALS is capable of providing valuable information concerning the type, size, and concentration of the defects or voids that exist within the material at the time of the annihilation process.

Coincidence Doppler Broadening Spectroscopy (CDBS) acts as a complementary technique, designed to measure the energy of gamma quanta released by positron-electron annihilation. The annihilation of a positron with an electron results in the production of two gamma photons. The energy of the 511 keV annihilation gamma qaunta is Doppler shifted due to the momentum of the annihilating electrons; the momentum of the thermalized positron can be neglected.

The shape of the Doppler broadened 511 keV annihialation photo peak carries information regarding the chemical environment surrounding the annihilation site, which is influenced by the momentum distribution of the electrons participating in the annihilation process. Analyzing the Doppler broadening spectra yields insights into the local electronic structure, chemical composition, and defect characteristics of the material\cite{PALS_review_1, PALS_review_2}.



\subsection{Experimental section}\label{experimentalpart} 

Defective LTO or "blue"-LTO was prepared using the pristine, commercial, LTO from S\"{u}dChemie (now Clariant). The blue LTO was obtained by annealing the pristine LTO, or white LTO, in an Ar/4\% \ch{H2} atmosphere, varying the holding time at 700-750 \degree C from 0 min up to 8h. Thermogravimetric analysys TGA/DSC1 STARe SYSTEM (Mettler Toledo, Switzerland), which is combined with a mass spectrometer (MS) as gas analysis system (Pfeiffer Vacuum, Germany) was used to confirm that oxygen was removed from the pristine LTO during the annealing treatment.\ref{fig:TGA-MS_index}
The structures of the pristine and the representative blue LTO, were identified using X-ray diffraction (XRD) analysis performed with a STOE STADIP diffractometer (STOE, Germany). The XRD pattern were recorded with \ch{Mo} $K_{\alpha_1}$ radiation ($\lambda$ = 0.70932 \AA ~ 50 keV, 40 mA) and a Mythen 1K detector. The diffraction patterns were collected with a capillary (0.3 mm) in a Debye-Scherer mode with a step size of 0.015\degree and measurement time of 7 s/step for overnight measurements.
The ICCD library was used to confirm the phases.\cite{GatesRector2019}
Rietveld refinement treatment of the data was performed with the Fullprof software package.~\cite{RodrguezCarvajal1993} 

The PALS\cite{Sperr2008} and CDBS\cite{Gigl2017} were performed at the high intensity positron beam for the Neutron induced Positron source Munich (NEPOMUC)\cite{Hugenschmidt2008} at the research neutron source Heinz Maier-Leibnitz (FRM II) of the Technical University of Munich.~\cite{Hugenschmidt2015,Gigl2017} 

\subsection{Theoretical Section}\label{Theoreticalpart}
Perdew-Burke-Ernzerhof (PBE) functional with a Hubbard U correction for computations at the semi-local Generalized Gradient Approximation (GGA) level of theory were used to accurately capture localized electrons (polarons),. All structural relaxations were executed using DFT+U, utilizing the planewave pseudopotential VASP code. Based on previous experimental measurements and DFT studies, we set the U value at 4.2 eV, a choice determined by reaching a polaronic state position ~1.0 eV below the conduction band minimum (CBM). Frozen-core projector-augmented wave potentials served as representations for ionic cores. To describe the electron-positron interaction and compute the lifetime of the positrons in our structures of interest, we employed the TCDFT implemented in the Abinit code\cite{TCDFT}. In the Abinit calculations, the same U value (4.2 eV) is used to consistently describe the relative positions of the polaronic state and the CBM. For the electron-positron correlation, both GGA and Local Density Approximation LDA types were tested on the widely used spinel-LTO bulk model\cite{Positron_Electron_GGA, Positron_Electron_LDA}. Based on the reference rutile/anatase \ch{TiO2} positron lifetimes data\cite{TiO2_lifetime}, we found that GGA tends to overestimate the positron lifetimes in oxides. For localized positrons found in surface and defect models, we found that only GGA can describe the correct positron density. Details of the comparison of positron lifetimes computed by these two methods can be found in \autoref{tab:ixcpositron_table}. To compute the S-parameters, we applied a Gaussian convolution to the spectral data, using a full-width at half-maximum (FWHM) of 3.5 mrad to emulate the experimental resolution observed in the reference study. Subsequently, we computed the S parameters over a range of 0 to 4.0 mrad.
For all computations, we set the plane wave cutoff energy at 500 eV. Structural relaxations were carried out until the force on any atom did not exceed 0.05 eV/Å. Details of different models (cell size, number of atoms, k-grid size) are tabulated in \ref{tab:model_info}. To maintain the neutrality of the unit cell, all calculations of charged defects (\ch{V^{+1}_{Li}}, \ch{V^{+2}_{O}}) incorporated uniform background compensation. 

The structural complexity, resulting in numerous LTO bulk/surface models, can be attributed to several factors, including the disorder brought about by \ch{Li^{16d}}/\ch{Ti^{16d}} fractional occupations, potential surface facets, and vacancy positions. For studying polaron stability on LTO surface, we built three LTO surface models with and without one oxygen vacancy. For studying positron lifetimes in LTO, we examine pristine bulk and surface LTO and bulk models with single-atomic vacancy(Li, Ti, and O).
Our initial step involves the construction of pristine bulk models, which will serve as the foundation for constructing surface models. The bulk models were first screened considering \ch{Li^{16d}} disorder by the well-developed force field\cite{LTO_force_field}. The relatively stable structures were then examined again using DFT~\ref{fig:bulk formation energy}. Three most stable bulk models in the corresponding series were used for constructing (111), (110), and (100) surface models, following the major low-index peaks observed in XRD experiments. To construct computationally feasible surface models with and without oxygen vacancies, we set a few constraints while limiting the number of models. We only considered stoichiometric and z-axis symmetric pristine surface models. The introduction of vacancies into the models, combined with the inequivalence of chemical environments, results in unique vacancy sites. These vacancies generate diverse environments conducive to potential polaron formation. Exploring all possible combinations is computationally intensive and may not be necessary when our primary goal is to capture the most pivotal information and relationships between these elements. Consequently, we set a fixed position for the oxygen vacancy in each surface series and inspect a restricted polaron distribution, aiming to identify trends within the same series and understand the disparities between the facets. The oxygen vacancy positions were hand-picked and located at the second layer in all three surface models.
\section{Results and discussion}
\subsection{Surface polaron}

\begin{figure*}[h]
\includegraphics[width = 1\textwidth]{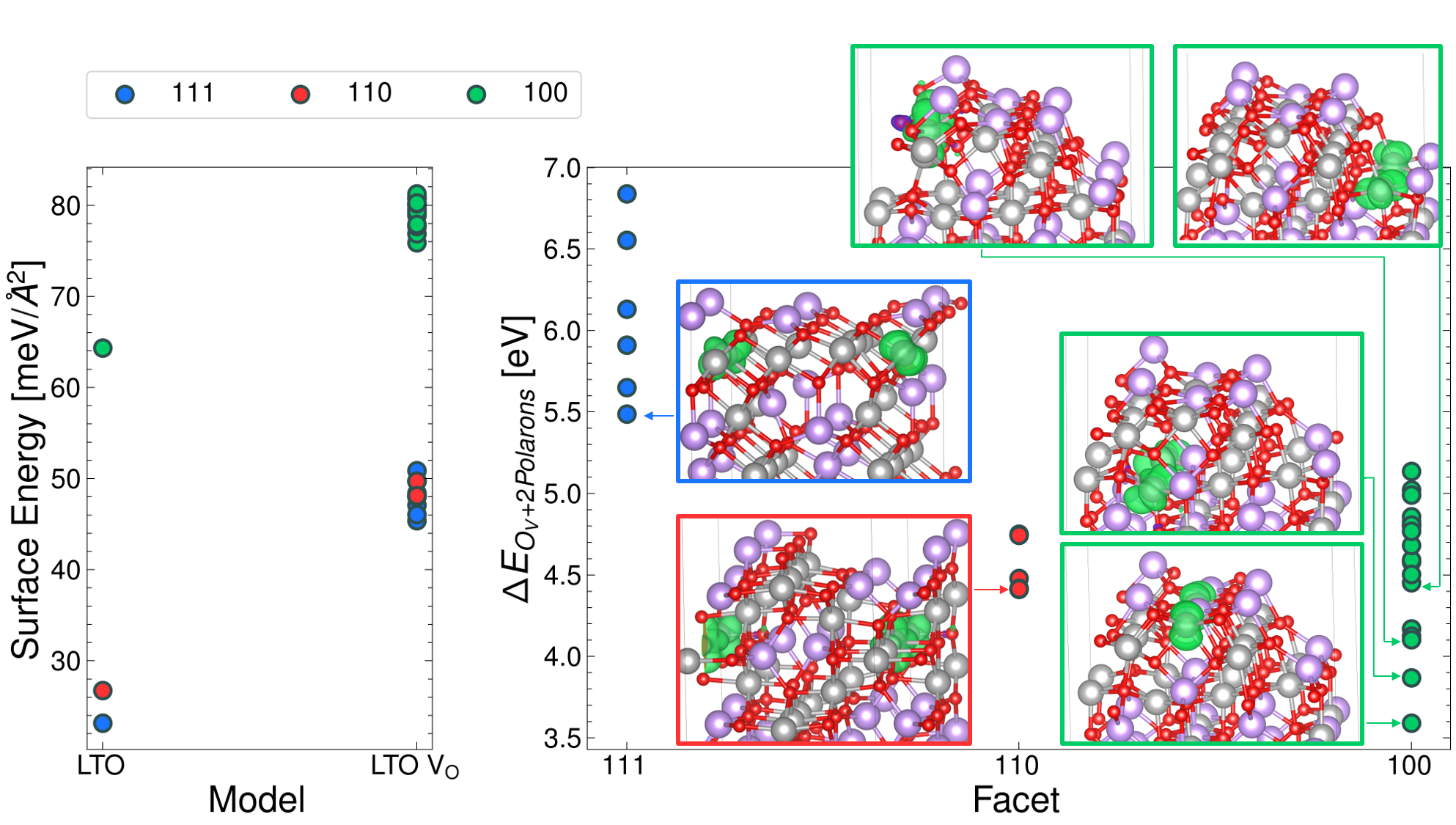} 
\caption{Surface energy of pristine models units(left) and energy cost of forming one oxygen vacancy and two polarons in surface models and the corresponding polaron density (right). Isosurface level of 0.01 e \AA$^{-3}$.}
\label{fig:LTO Ov surface formation energy}
\end{figure*}

The \ch{V_{O}}-induced polaron and its hopping in LTO has been considered as one of the roots of enhanced conductivity in blue LTO\cite{Kick2020}. To investigate the origin of the facet-dependent conductivity difference, we examined the stability of polarons on different LTO surface models with various polaron distributions. During the reduction process in blue LTO experiments, oxygen vacancies and associated polarons are generated in LTO. The energy cost for this process consists of two terms: oxygen vacancy formation and polaron formation. 

\begin{equation}
\begin{split}
\Delta E_{O_v \mathbin{+} 2 \text{Polarons}} &= E_{O_v \mathbin{+} 2 \text{Polarons}} + \frac{1}{2} E_{O_2} \\
&\quad - E_{\text{surface}}
\end{split}
\end{equation}

We found a significant variation in the energy cost among the three facet series due to their structural differences. The polarons (\ch{Ti^3+}) tend to elongate the Ti-O bonds and distort the octahedral (\ch{TiO6}) geometry, which results in the lowest energy cost for the (100) series due to the built-in distortion, as illustrated in \ref{fig:LTO Ov surface formation energy}. 

The \ch{TiO6} motifs on the (100) surface are highly distorted and provide a suitable environment for polaron formation. This is due to both the existing distortion and the less rigid structure resulting from the disorder. Compared to the (111) surface, which is the most stable and widely studied, the (100) surface has a lower energy cost of polaron formation, ranging from 3.59 to 5.14 eV, while the (111) surface has a minimum energy cost of 5.48 eV. Given that in the (100) surface the self-interactions between the polarons and their periodic images may be stronger, which arises from the smaller x-y lattice, the polaron formation energy cost is even slightly overestimated. The structural disorder on the (100) surface creates various \ch{Ti} sites that can host polarons, including the \ch{TiO5} motifs that are generated by the \ch{V_{O}}. Similar \ch{TiO5} motifs cannot form polarons in the (111) and (110) surface models due to their rigidity and the large structural change caused by the adjacent \ch{V_{O}} and polaron. In addition to the facet-dependent polaron formation tendency, polarons generally prefer to form on or near the surface, which is consistent with the previous findings on the (111) surface\cite{Matthias_111_JCP}. For comparison, in the bulk region, the polaron distribution is mainly influenced by the \ch{V_{O}} due to their strong attraction\cite{Matthias_JPCL}.

The polaron stability result shows that polarons form more easily on the (100) surfaces of LTO compared to the (111) surfaces. It is established that polaron formation can augment both electronic and ionic conductivity in LTO, facilitated by polaron hopping and polaron-assisted \ch{Li^+} diffusion\cite{Matthias_JPCL, Matthias_polaron_assisted}. This explains why LTO with mixed (100)/(111) surfaces performs better than LTO with exclusively  (111) surfaces in experiments\cite{Pal_111vs100}. 
Furthermore, such a polaron-induced surface stability trend is in line with the comparison between different synthesis strategies. The synthesis method affects the surface stability of LTO, which is related to the degree of polaron formation induced by \ch{V_{O}}. The XRD data of LTO prepared by carbothermal reduction\cite{carbothermal_reduction} manifest a prominent (100) peak with an increment in carbon black content. Carbothermal reduction introduces \ch{V_{O}} in \ch{TiO2} before reacting with \ch{Li2CO3} to form LTO. In contrast, the XRD patterns of white and blue LTO (pre and post \ch{H2} treatment) do not exhibit noticeable disparities, given that the oxygen vacancies are introduced at later stages\cite{Blue_LTO_Li_diffusivity, blue_color_fading}.

\subsection{Positron lifetimes}
We used positron annihilation lifetime spectroscopy (PALS) with two kinetic energies (1 keV and 18 keV) to investigate how \ch{H2} treatment affects the structure and defects of LTO. The 1 keV positrons probe the surface region, while the 18 keV positrons probe the bulk region. Table \ref{tab:exp_positron_lifetimes} shows the two main positron lifetimes and their intensities for white and blue LTO. To identify the defect types and structures associated with these lifetimes, we compared them with theoretical lifetimes calculated by two-component DFT (TCDFT). We first examine three reference materials, \ch{BCC-Li}, \ch{Rutile-TiO2} and \ch{Anatase-TiO2} to see whether the TCDFT is able to describe the positron state within the model well and give the correct positron lifetimes compared to the values reported experimentally. The result of \ch{BCC-Li}, 297 (291 $\pm$ 6) ps \cite{Li_exp_PL}, indicates that the method we used is good enough to compute theoretical positron lifetimes, which matches the experimental(in parenthesis) value well. For \ch{TiO2}, both \ch{Rutile-TiO2}, 157 (135 and 148 $\pm$ 4) ps, and \ch{Anatase-TiO2}, 191 (~200) ps have a good agreement with the experiments\cite{TiO2_lifetime, TiO2_exp_PL}. More details regarding the benchmark are discussed in Table \ref{tab:ixcpositron_table}.
\subsubsection{Surface region (1 keV)}

\begin{figure}[h]
\includegraphics[width =\columnwidth]{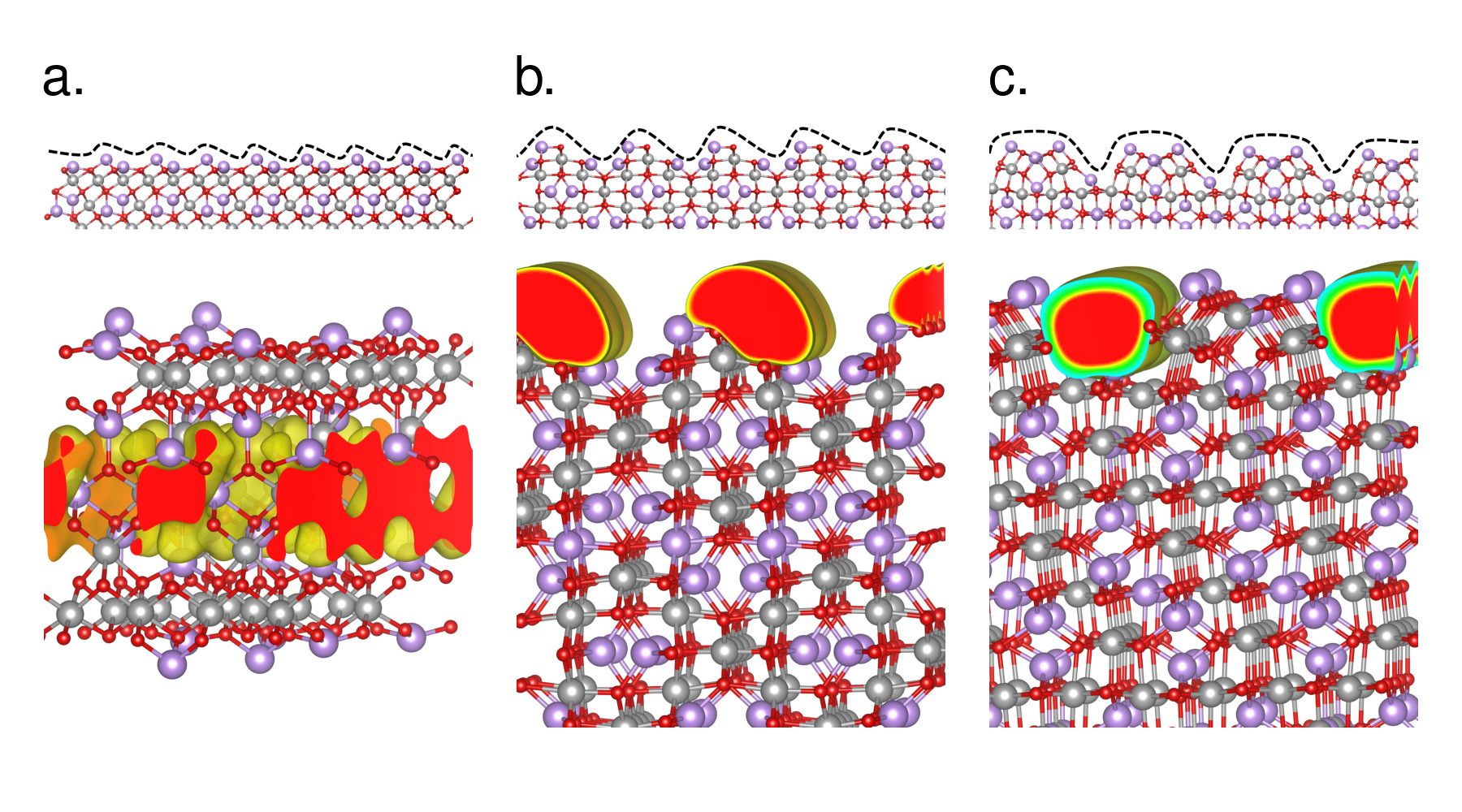} 
\caption{Positron density and  surface roughness in a). (111), b). (110) and c). (100) surface models. Isosurface level of 0.0015 e \AA$^{-3}$.}
\label{fig:Positron_density_surface}
\end{figure}

Two main lifetimes observed in the white LTO surface region are 198 $\pm$ 2 ps (48.0 $\pm$ 1.1\%) and 377 $\pm$ 2 ps (51.9 $\pm$ 1.1\%). We first compare them to the theoretical values from surface models. Three surface models, (111), (110) and (100), were created and used for computing positron lifetimes. In (100) and (110) models, positron densities are distributed on the surface, giving lifetimes of 379 and 412 ps, respectively. The experimentally observed 377 ps likely originates from positrons trapped on the surface, as indicated by the positron distribution in the (100) and (110) models, which result in longer lifetimes. However, in the (111) model, positron density distributes in the middle of the slab (i.e., in the sub-surface region) rather than on the surface. This might be due to the fact that we have an ideal flat surface structure in the (111) model that makes it hard to trap positrons. Meanwhile, the rough surface structure of (100) and (110) both provide "ditch-like" trapping sites that stabilize positrons well, see \ref{fig:Positron_density_surface}. In other words, when being injected into the (111) region, a positron should behave more like when injected into bulk. Indeed the 198 ps one is close to the theoretical positron lifetime in bulk(defect-free), which is 186 $\pm$ 0($\tau_b$)  ps averaged from 27 stable models \ref{fig:bulk formation energy}. Moreover, the positron densities in (111) surface and all bulk models both tend to distribute along the 16c site network and have a denser population around \ch{Li^{16d}} sites, which should provide less repulsion compared to \ch{{Ti}^{16d}}.

After treatment with \ch{H2}, the blue LTO surface region produced two distinct positron lifetime signals, 224 $\pm$ 2 ps (62.0 $\pm$ 2.0\%) and 362 $\pm$ 4 ps (37.9 $\pm$ 2.0\%), which differ from the white counterpart. The longer lifetime is still close to the (100) theoretical value but is present in smaller proportions after reduction. The decrease in intensity of (100) may be due to competition from the species responsible for the new positron lifetime signal of 224 ps. This value is much higher than theoretical $\tau_b$ (186 ps) and is quite similar to the values we computed for oxygen vacancy with +2 charge(\ch{V_{O}^{2+}}) models (224 $\pm$ 1) ps. In these models, one oxygen vacancy is introduced and two electrons are removed to form \ch{V_{O}^{2+}}, which do not have polarons around the oxygen vacancy site. We found that in these models, positron densities distribute similarly like in bulk (along the 16c site network) but only have a small portion of density go into the vacancy site. Consequently, the computed positron lifetime (224 $\pm$ 1) ps is slightly higher than $\tau_b$. One possible reason may be that the cations around the oxygen vacancy site have positive charges, which repel positrons and make the anion vacancy hard to trap the positron fully. The same calculations were carried out without the +U scheme and gave a similar result ($\sim$ 226 ps). It is worth noting that the lifetimes we discussed so far are computed using the geometry relaxed without positron-induced forces. The rule of thumb for considering the positron-induced geometry change is that only the localized positron alters the local structure since the delocalized one only has a vanishing density. In principle, the highly localized trapped positron at void-like defects is analogous to the self-trapping small polarons in ionic crystals. The localized positrons can affect their surroundings, altering the electron density distribution and the local structure near the trapped positron. However, what we saw in the \ch{V_{O}^{2+}} is that only a portion of positrons goes inside the vacancy. Furthermore, throughout the self-consistent positron-electron scheme at a fixed structure, the computed lifetimes at each step actually stay almost the same (218 - 224 ps), meaning the delocalized positron does not alter the electron density. The highly distributed positron density hints that this kind of anion vacancy, surrounded by cations, makes it hard to trap positrons. However, when the anion vacancies have polaron nearby, the interaction between these two species with opposite charges makes things much more appealing. The \ch{V_{O}^{0}} model with two polarons nearby can effectively trap the entire positron resulting in notable interaction between positron and electron that alters the electron density resulting in a longer lifetime, which we will discuss further in the later section.

\begin{table*}
  \caption{Experimental positron lifetimes measured for different samples. $\tau_1$ and $\tau_2$ represent the two major lifetime components ,while $I_1$ and $I_2$ are their corresponding intensities}
  \label{tab:exp_positron_lifetimes}
  \centering
  \begin{tabularx}{\textwidth}{lXXXX}
    \hline
    Sample & $\tau_1$ (ps)  & $I_1$(\%) & $\tau_2$ (ps)  & $I_2$(\%)  \\
    \hline
    W-LTO-Bulk     & 141 $\pm$ 4  & 32.7 $\pm$ 2.2  & 295 $\pm$ 6 & 61.4  $\pm$ 1.5  \\
    B-LTO-Bulk     &  83 $\pm$ 7  &  6.7 $\pm$ 0.5  & 254 $\pm$ 1 & 91.2  $\pm$ 0.4  \\
    W-LTO-Surface  & 198 $\pm$ 2  & 48.0 $\pm$ 1.1  & 377 $\pm$ 2 & 51.9  $\pm$ 1.1  \\
    B-LTO-Surface  & 224 $\pm$ 2  & 62.0 $\pm$ 2.0  & 362 $\pm$ 4 & 37.9  $\pm$ 2.0  \\
    \hline
  \end{tabularx}
\end{table*}

\begin{figure}[h]
\includegraphics[width =\columnwidth]{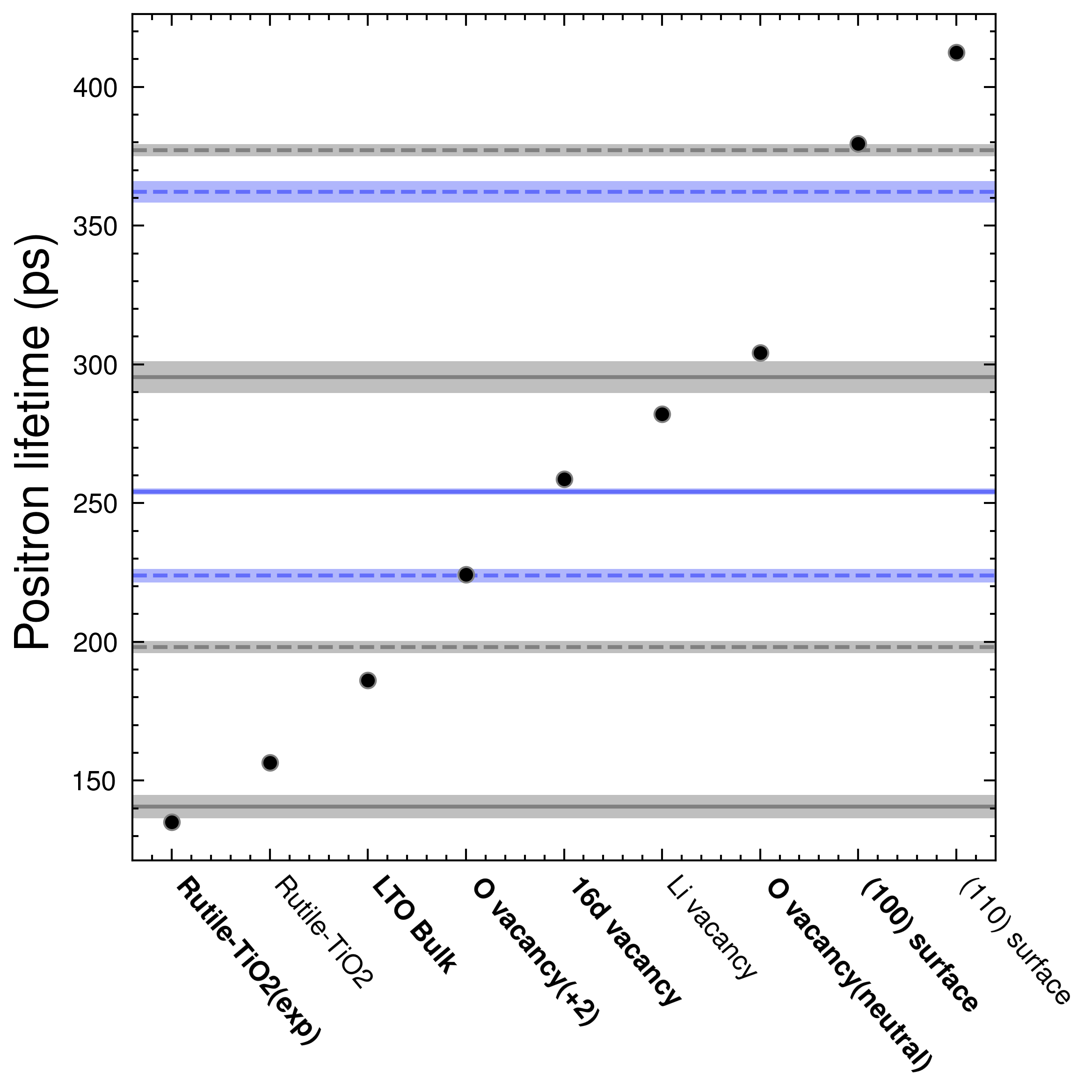} 
\caption{Experimental (lines) and theoretical (dots) positron lifetimes. Data from white and blue LTO are shown in grey and blue respectively, where the dashed lines come from 1 keV and solid lines come from 18 keV results.}
\label{fig:Positron lifetimes}
\end{figure}

\subsubsection{Bulk region (18 keV)}
\begin{figure*}[h]
\centering
\includegraphics[width=\textwidth,keepaspectratio]{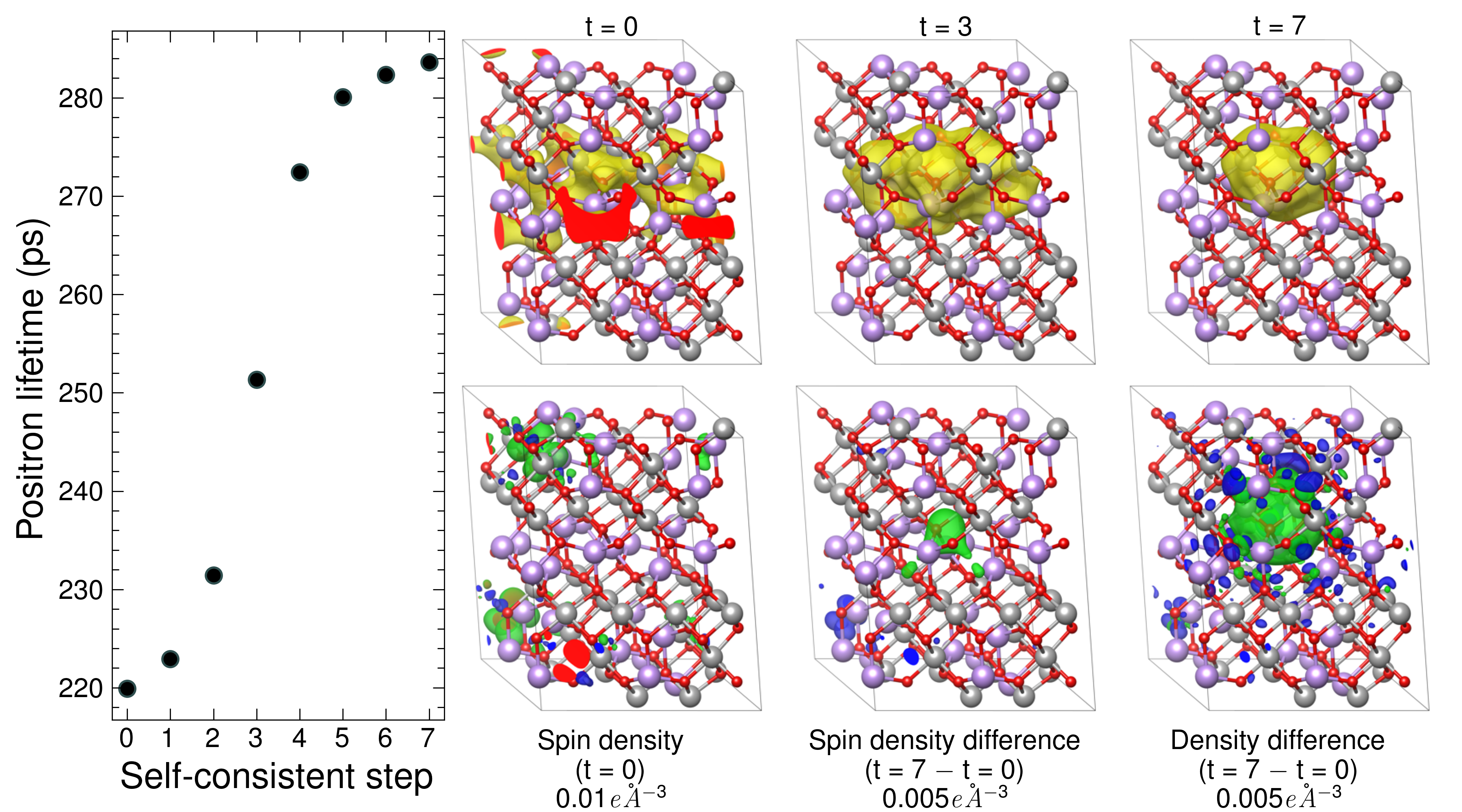}
\caption{Left: positron lifetimes along the self-consistent steps (t = 0 to 7). Right: positron density distributions (upper) and spin-density and density difference (bottom) along self-consistent steps. Yellow and green/blue represent positron and electron density, respectively. Isosurface level of 0.0015 e \AA$^{-3}$ is used in positron density plots. Different isosurface levels were adopted for electron density as shown in the figure to have a clear view.}
\label{fig:Positron_lifetimes_density_changers_Ov}
\end{figure*}

In the bulk region of LTO, two main positron lifetimes of 141 $\pm$ 4 (32.7 $\pm$ 2.2\%) and 295 $\pm$ 6 (61.4 $\pm$ 1.5\%) ps were observed. These values differ significantly from the theoretical bulk value($\tau_b$, $\sim$186 ps). Comparing them with theoretical positron lifetimes computed using TCDFT and experimental values reported in the literature revealed that the 141 ps lifetime likely originates from rutile \ch{TiO2} impurities, a common precursor and impurity in LTO. The reported positron lifetimes of rutile \ch{TiO2} are about 135 ps \cite{TiO2_lifetime} and 148 $\pm$ 4 ps\cite{TiO2_exp_PL}, and the lifetime computed in this study is 157 ps. This suggests some \ch{TiO2} domains are present in the LTO sample. Once injected into these domains or nearby regions, a positron can diffuse within this domain and give the 141 ps signal. Our XRD data also indicates the existence of rutile/anatase \ch{TiO2} in the sample as an impurity from the pristine sample. 
The 295 ps lifetime is much longer than $\tau_b$ and lower than the ones associated with surface models. Positron lifetimes higher than $\tau_b$ by approximately 100 ps are typically interpreted as signals from vacancy clusters or grain boundaries, which are generally considered more difficult to form and only present in smaller proportions. However, this signal covers about 61.4\% intensity, which indicates it is likely to be abundant. Compared to the theoretical results, it turns out that this 295 ps lifetime  originates from an oxygen vacancy with polarons around it, i.e., \ch{V_{O}^{0}}. Unlike the \ch{V_{O}^{2+}}, we found that positrons can be trapped well in the oxygen vacancy site when polarons are nearby in the \ch{V_{O}^{0}} models. This effect can be understood by studying the positron/electron density change along the self-consistent calculation, shown in \ref{fig:Positron_lifetimes_density_changers_Ov}. In the first few steps of the self-consistent scheme at a fixed structure(relaxed without positron), the positron lifetimes are about 220-230 ps, which is close to the result of \ch{V_{O}^{2+}}. The positron densities in these steps are delocalized, like the ones in pristine bulk and \ch{V_{O}^{2+}}. However, interestingly, the positron gradually moves into the oxygen vacancy site along the process and, in the end, is completely trapped in the void. On the other hand, the spin-charge density shows that a part of the electron density moves from polarons to the void region. Consequently, a huge jump in the lifetime occurred from 212 to 284 ps after self-consistency was reached. This result sheds light on the long-lasting problem of how to "directly" detect oxygen vacancy in materials\cite{Jakes2015}. Here we are able to determine whether the oxygen vacancy exists and simultaneously have information about its surroundings by combining experimental PALS data and theoretical calculations. 
Marinopoulos previously studied on theoretical positron lifetimes in cubic yttria-stabilized zirconia(YSZ) defect models showing that the relaxation caused by the positron trapped in the vacancy might be important for some cases in terms of a change in the lifetime by ~20 ps\cite{YSZ_relaxation_under_present_of_positron}. Another earlier study also claimed that positron-induced lattice relaxation is indispensable for the existence of localized positron states\cite{Energetic_of_positron_at_vacancies_in_solid}.
After further relaxation under the presence of positron, the lifetime of positrons trapped in the \ch{V_{O}^{0}} increases a bit to 304.14 $\pm$ 1.12 ps, averaged from three distinguish polaron distributions. This suggests that there are some pre-existing oxygen vacancies accompanied by polarons around them in the bulk region of the white LTO sample.

In the blue LTO, two lifetimes previously found in the white LTO disappear and two new ones emerge. The majority of the detected annihilation events give a lifetime of 254 $\pm$ 1 ps that is corresponding to 91.2 $\pm$ 0.4\% of the signal intensity, which is close to the one computed from \ch{V_{Li/Ti}^{16d}}. Noted here, due to the material nature of LTO, the 16d trapping site could possibly created by \ch{V_{Li}^{16d}} and \ch{V_{Ti}^{16d}} due to the partial occupancy. 
The vacancy at the 16d site generally gives a lifetime of 259 $\pm$ 1 ps based on the TCDFT results. In theory, the bonds between Ti and the six coordinated oxygen atoms should be stronger than those between Li and oxygen, meaning that the vacancy at the 16d site likely comes from Li migrating out of the sites. Moreover, the \ch{V_{Li}^{16d}} at the Li-rich region is more stable than other possible sites by more than 0.4 eV based on our calculations. This is in line with a recent MD study on lithium diffusion, which pointed out that 16d positions act as trapping sites for Li vacancies\cite{Li_16d_trapping_site}. During the normal battery operation, \ch{V_{Li}^{16d}} are considered immobile due to the larger hopping barrier, which has been observed in NMR experiments\cite{Li_diffusion_MD, Immobile_16d_Li}, again pointing  out thatit is stable in LTO. Additionally, the \ch{Li^{16d}} at the Li-rich layer should be much more mobile than in other 16d sites due to fewer constraints  and repulsion from \ch{TiO6} motifs nearby. During the \ch{H2} treatment, samples were heated up to $700-750 \degree C$  which provided an opportunity for \ch{Li^{16d}} to move. Interestingly, another lifetime observed here, 83 $\pm$ 7 (6.7 $\pm$ 0.5\%), might hint at where the mobile Li-ions move. It has been found that \ch{Rutile-TiO2} can undergo lithiation and storage of Li-ions\cite{TiO2_lithiation}. It was also found that \ch{Rutile-TiO2} is able to provide a fast Li-ion diffusion channel and improve the electrochemical performance of LTO\cite{hwang_2019_polygonal}. On the other hand, polarons in \ch{TiO2} have been studied extensively and are considered essential in improving electronic conductivity, i.e., it is mobile. It would not be a surprise that when the samples are heated up, these two mobile charge carriers, with opposite charges, move together into the \ch{rutile-TiO2} impurity domain while keeping the local charge neutrality\cite{Schleker2023_2}. The theoretical lifetime we got from lithiated \ch{TiO2} is 95.01 ps, close to the observed value. Although the effect of the presence of \ch{V_{Li}^{16d}} is not clear so far, Zhan et al. did explore that the \ch{Li^{16d}} and \ch{Ti^{16d}} arrangement around the face-sharing \ch{Li^{8a}} and \ch{Li^{16c}} could introduce different degrees of local distortion that lower the Li hopping barrier(\ch{Li^{8a}} $\leftrightarrow$ \ch{Li^{16c}})\cite{Kinetic_pathway_of_Li}. Arguably, \ch{V_{Li}^{16d}}, which also breaks the symmetry, should be able to introduce local distortion and accelerate Li diffusion. 

Although the annihilation associated with 83 ps lifetime only contributed a small portion due to the high positron affinity of \ch{V_{Li}^{16d}}, this lithiation phenomenon is crucial for understanding the material nature and possible origin of the increasing electronic conductivity and Li diffusion rate both observed in blue LTO ($D_{Li}$ increases by a factor of 140) and LTO-\ch{TiO2} material (by a factor of 10)\cite{Blue_LTO_Li_diffusivity, LTO_TiO2_Li_diffusion}. The significant increases in $D_{Li}$ found in blue LTO might root from the creation of \ch{V_{Li}^{16d}} which can introduce local distortion to break \ch{Li^{8a}-O4} and \ch{Li^{16c}-O6} symmetry to potentially lower the effective coordination numbers\cite{Kinetic_pathway_of_Li}.
The whole picture of the structural changes derived from the experiments and theory in the bulk region before/after the \ch{H2} treatment obeys charge neutrality. The positive charge Li at 16d sites and negative charge polaron move together into \ch{TiO2} regions when being heated and leave the negative charge \ch{V_{Li}^{16d}} and positive charge \ch{V_{O}} in the LTO domains. The missing signal from \ch{V_{O}} is likely due to (i). the stronger positron affinity of \ch{V_{Li}^{16d}}, (ii). losing polarons make oxygen vacancy hard to trap positrons.

Notably, the positron lifetime of single atomic vacancies in elemental and compound semiconductors studied earlier typically is $\sim$30 ps longer than the defect-free one\cite{GaAs_positron_lifetimes, Si_positron_lifetimes}. However, in metal oxides, the different types of vacancy sites, such as cation and anion vacancies, can significantly impact the chemical environment. The vacancies at 8a, 16d and 32e sites in LTO provide a versatile environment for positrons, leading to unique lifetimes we observed and computed.

\subsection{Local environment through CDBS}
\begin{figure}[h]
\includegraphics[width =\columnwidth]{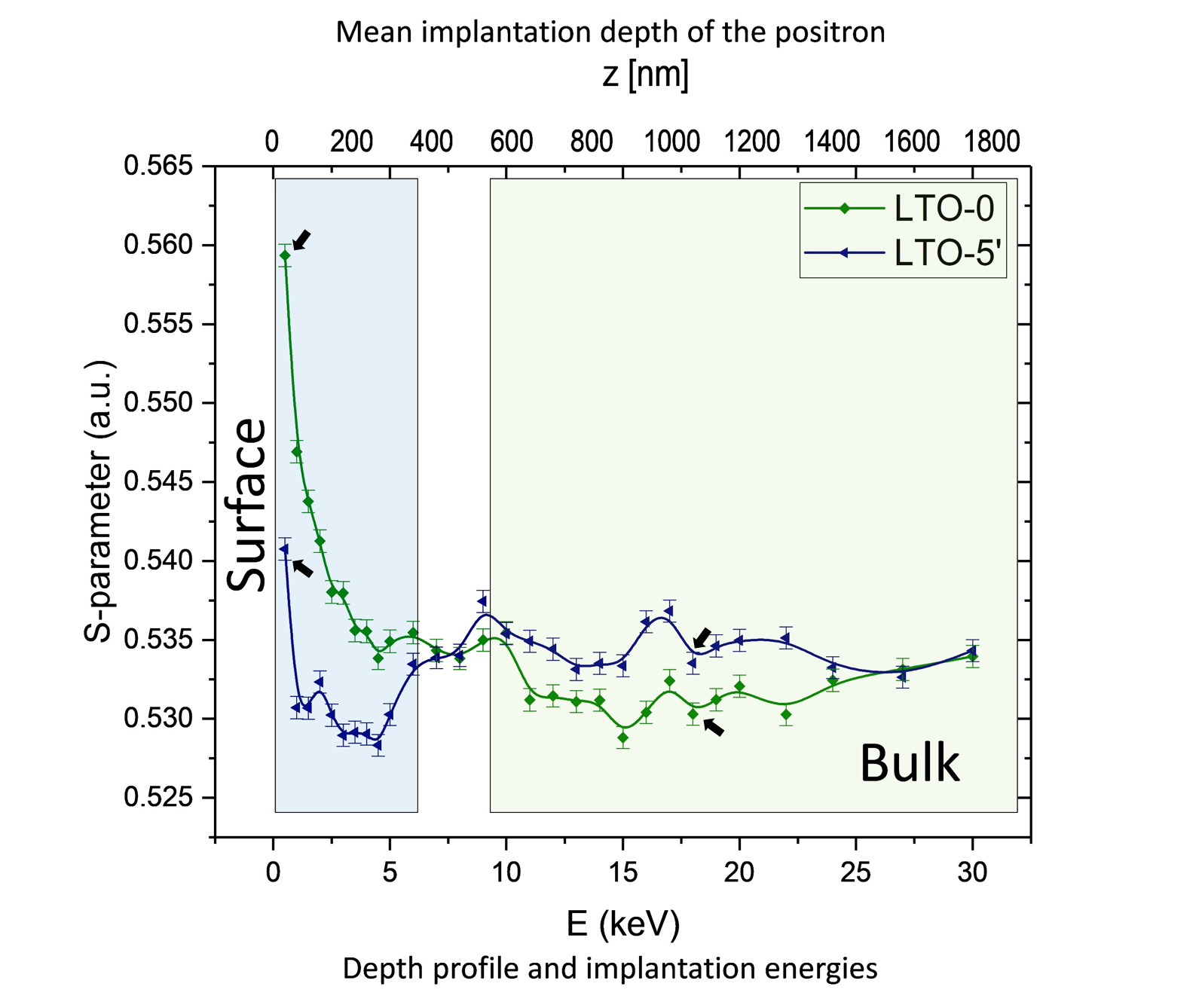} 
\caption{Experimental CDBS depth profile of W/B-LTO. The areas highlighted in blue and green show two distinguished signals, representing surface and bulk regions. The arrows point out the sample for measuring positron lifetimes(1 keV and 18 keV)}
\label{fig:CDBS}
\end{figure}

To further investigate the local structure at the sites of annihilation events and the distribution of the defects, we conduct CDBS measurements using positrons with various kinetic energy, exploring different regions of the samples. These provide two parameters, shape (S) and wing (W), linked to annihilation events with different electron momentum distributions. 
The S-parameters computed from the Doppler-broadened spectrum are shown in \ref{fig:CDBS}. Two distinguished regions representing surface and bulk are found and highlighted. The higher S-parameter comes from more low-momentum electrons participating in the annihilation events, which usually correlate to an increased defect density or a larger defect size. 
\subsubsection{Surface region}
In the surface region (low kinetic energy), relatively high S-parameters are recorded in both white and blue LTO. This aligns with the species we assigned to the 1keV signal, surface (100), where the positron trapping in a long ditch-like void. The drop observed from the 1 keV points of white and blue LTO is also consistent with the (100) signal intensity drop found in PALS, shown in \ref{tab:exp_positron_lifetimes}. Moreover, the S-parameter decreases across the surface regions, resulting from less annihilation happening on the surface. In the blue LTO surface region, the S-parameter converges and forms a small plateau about 0.528 within 2.5 - 5 keV. This matches with the theoretical S-parameter computed from \ch{V_{O}^{2+}}, 0.528, which is slightly higher than the one from bulk, 0.522. In other words, the (100) surface signal no longer exists when the injected positron goes to the deeper area. Most of the positron annihilation happens in \ch{V_{O}^{2+}}, which confirms that the \ch{H2} treatment introduces a considerable amount of oxygen vacancy in the subsurface region, in line with PALS's assignment. 

\begin{figure}[h]
\includegraphics[width =\columnwidth]{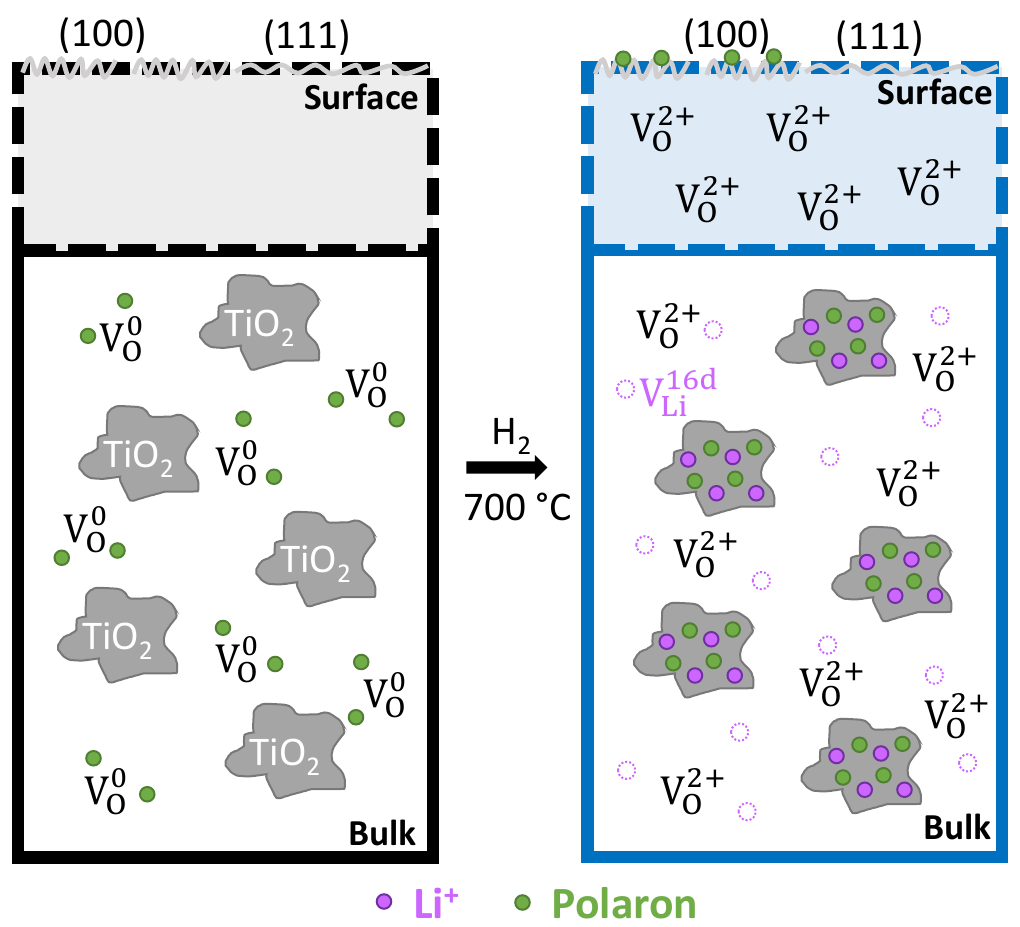} 
\caption{Schematic representation of the distribution of ions and defects in W/B-LTO before and after \ch{H2} treatment derived from combining PALS, CDBS and TCDFT results.}
\label{fig:defect_distribution}
\end{figure}

\subsubsection{Bulk region} 
In the bulk region, the S-parameters difference between white and blue LTO is small, although the species found by PALS data are completely different. The reason behind this is that the S-parameter computed from the measured spectrum is a statistical result that comes from all the annihilation events from different defect types. Even though the defects found in white and blue LTO are different, the overall S-parameter is found to be close. We can estimate the experimental S-parameter using the weighted sum of the theoretical ones based on the signal intensities measured in PALS. In the white LTO bulk region, it consists of 32.7\% \ch{TiO2}(S=0.512) and 61.4 \% \ch{V_{O}^{0}}(S=0.531), which ammount to a S-parameter of about 0.523. On the other hand, in the blue LTO, 6.7\% \ch{lithiated-TiO2} and 91.2\% \ch{V_{Li}^{16d}}(S=0.55) are found, leading to a S-parameter about 0.536. We use the same theoretical parameter for \ch{lithiated-TiO2} as for \ch{TiO2}, which should be an acceptable estimation considering its small fraction in the sample. These values roughly represents the exact and relative positions at 18 keV in \ref{fig:CDBS}. The fluctuating lines in the bulk region could be caused by the impurity \ch{TiO2} domain distributing unevenly. The creation of \ch{V_{O}^{2+}} from \ch{H2} treatment only occurs in the surface region, which can explain the blue color fading after five weeks found in a previous study\cite{blue_color_fading}. The depth profile derived from PALS, CDBS and TCDFT results gives us a clear picture of the defect distribution before and after \ch{H2} treatment shown schematically in \ref{fig:defect_distribution}. This method is generally applicable ito study defect engineering in complex materials for designing and refining the process.

\section{Conclusion}\label{conclusion}

In this work we tailored the formation of oxygen defects on the spinel structure of LTO. We observed the 
distribution of the defects from surface to bulk via positron annihilation spectroscopy. Additionally the use of DFT+U and TCDFT allowed to deconvolute the presence and the formation of the polarons within the pristine and the blue-LTO. The simulations were crucial to reveal that the (100) exposed surface accommodates more charge carriers (polarons) due to its inherently lower formation energy.  
We showed that the integration of first principle calculations with DFT+U, TCDFT for the interpretation of experimental techniques using positron as source represents a promising approach to study defect structure in complex materials.

\begin{acknowledgement}



The authors thank Hubert Gasteiger, Domink G. Haering and Karin Kleiner for valuable contribution to the experimental aspects of this project. 
For the valuable discussion on the PAS and for fruitful contribution to the materials synthesis and preparation
P. Philipp M. Schleker, Simone S. Kocher and Svitlana Taranenko.
The authors gratefully acknowledge the computational and data resources provided by the Leibniz Supercomputing Centre (LRZ) and Julich.

\end{acknowledgement}






\bibliography{LTO_Positron.bib}

\end{document}


\section{Experimental details}
\subsection{Sample preparation and characterization}
\subsubsection{Thermogravimetric analysys with a mass spectrometer (TGA-MS) for the sample preparation} \label{TGA-MS}

\begin{figure}[h]
\includegraphics[scale=.15]{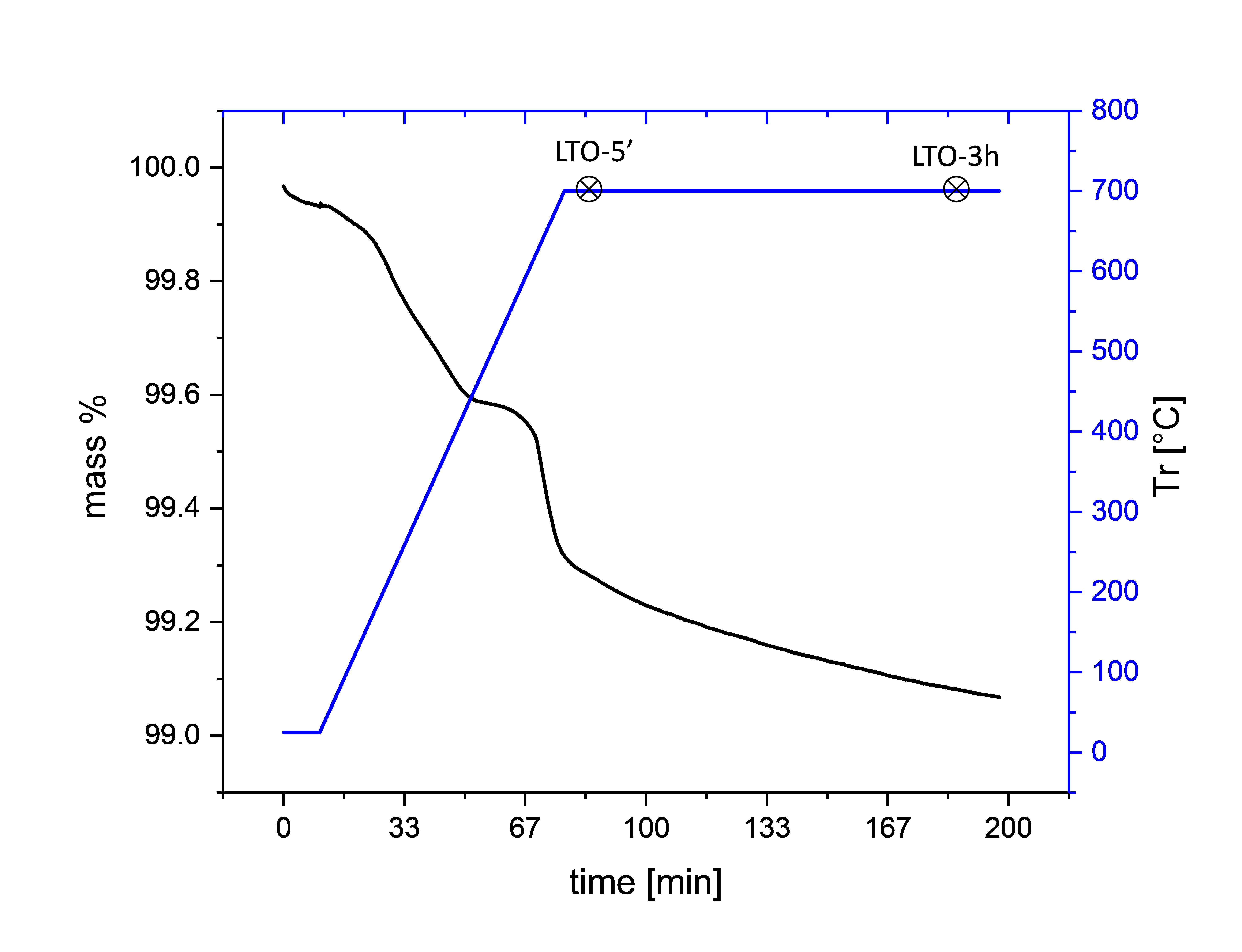} 
\caption{Thermogravimetric analysis for the pristine \ch{Li4Ti5O12}. LTO-5' and LTO-3h indicates two examples with the variation of the holding time, once the annealing temperature of 700 \textdegree C was reached.}
\label{fig:TGA-MS_index}
\end{figure}

In order to ensure the synthesis of \ch{Li4Ti5O12} LTO with oxygen defects, a thermogravimetric analysys with a mass spectrometer (TGA/MS) instrument, was used to monitor the annealing treatment. 
All the measurements were performed with a TGA/DSC1 STARe SYSTEM (Mettler Toledo, Switzerland), which is combined with a mass spectrometer (MS) as gas analysis system (Pfeiffer Vacuum, Germany). The synthesis were performed in the temperature range of 25 \textdegree C - 700/750 \textdegree C using sapphire crucibles (Ø 12 mm) in \ch{Ar}/\ch{H2} 5\% atmosphere at a heating rate of 5 K/min. The TGA connected with the MS gas analysis allows us to identify the components released during the heating treatment (in a classic way the decomposition of the investigated samples). Hence, we can confirm the depletion of the oxygen from our pristine LTO. 
We create a series of defective  or the so called"blue" LTO  starting from the commercial "white" LTO (S\"ud Chemie, now Clariant). We prepared the blue LTO by varying the time holding at the annealing temperature 700-750 \textdegree C from five minutes to eight hours ~\ref{fig:TGA-MS_index}. The intensity of blue theoretically corresponds to a different oxygen concentration.

\begin{figure}[h]
\includegraphics[scale=.5]{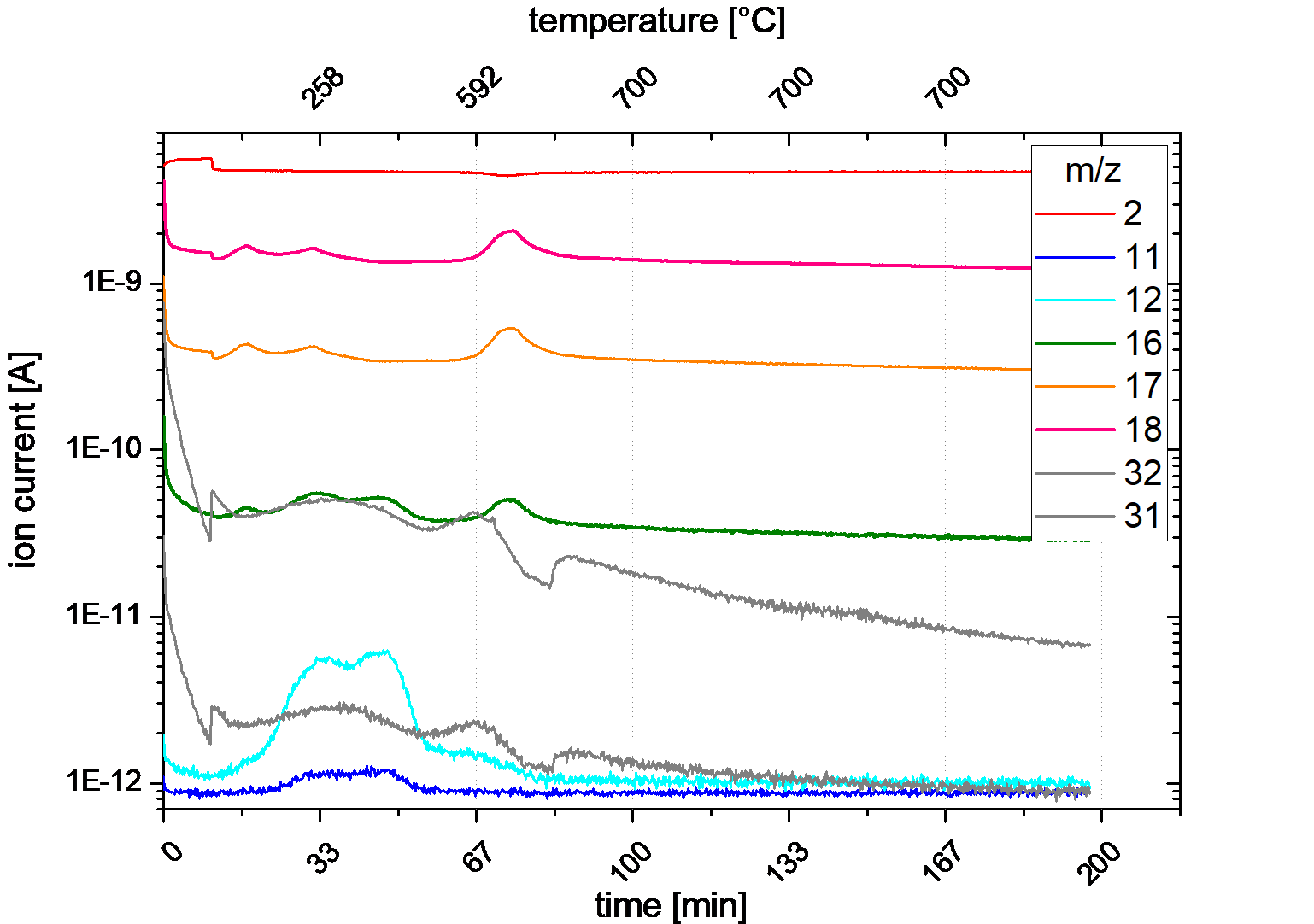} 
\caption{Thermogravimetric analysis and mass spectrometric measurements, while performing the annealing protocol for tailoring the oxygen vacancy formation in \ch{Li4Ti5O12} in an Ar/5 \% \ch{H2} atmosphere. The decrease in the signal at m/z 32 (\ch{O2}), with the raise in m/z 18 (\ch{H2O}) within the temperature range of 592 till 700 \textdegree C is an indication of the oxygen depletion from the pristine \ch{Li4Ti5O12}.}
\label{MScomponents}
\end{figure}

\FloatBarrier

\subsection{Samples Characterization}
\subsubsection{Powder X-ray measurements and the Rietveld refinement strategy}

Rietveld Refinement was done with FullProf v7.95 (Jan 2023). All phases were refined with Thomson-Cox-Hastings profile functions convoluted with an asymmetry correction according to the Finger-Cox-Jephcoat approximation as implemented in FullProf. Asymmetry parameters were constrained for all phases. Background was refined with a Chebychev-Polynomial with 10 Parameters.

\begin{figure}
\includegraphics[scale=1.0]{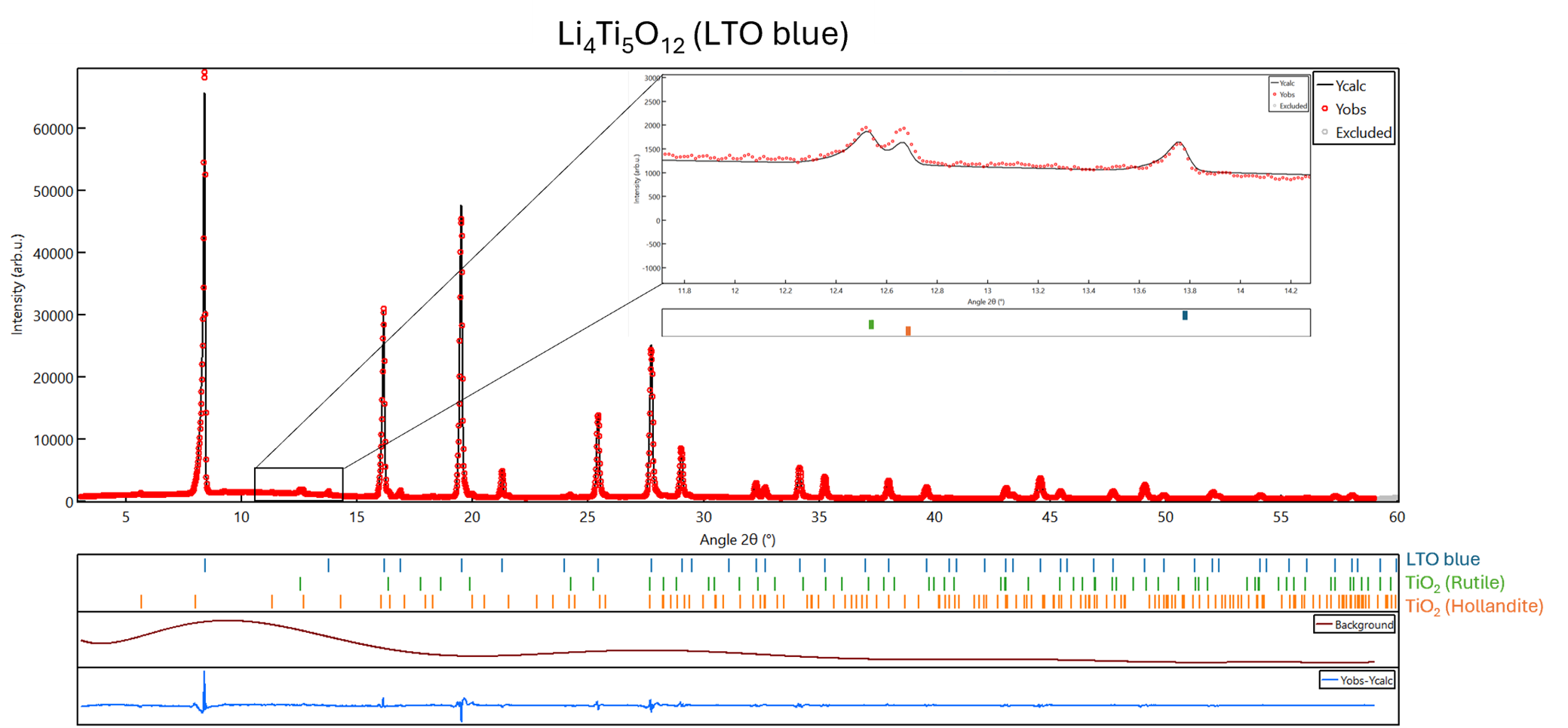} 
\caption{Rietveld refinement of Li$_4$Ti$_5$O$_{12}$ (LTO blue) of the X-ray powder diffraction pattern using three phase for refinement. Main phase LTO blue. Impurity phases TiO$_2$  rutile-type and TiO$_2$ in an empty hollandite-type.The XRD pattern were recorded with \ch{Mo} $K_{\alpha_1}$ radiation ($\lambda$ = 0.70932 \AA, \~ 50 $keV$, 40 $mA$)}
\label{BlueLTO}
\end{figure}

Initial structure models for refinement: \\
LTO blue: \ch{Li4Ti5O12} (ICSD-\#160655) \\
\ch{TiO2} rutile-type (ICSD-\#121636) \\
\ch{TiO2} hollandite-type (ICSD-\#120242) \\

Refinement results: \\
$R_{exp} = 2.74, R_{wp} = 0.0682, Chi^{2} = 6.35 $\\
$R_{Bragg}$ LTO blue = 0.021 \\

Phase 1 $:$ LTO blue (mass fraction $\approx 98(1)$ $\%$ ) \\ 

Composition from refinement $:$ ${Li_{1.43(11)}Ti_{1.50}{2(45)}O_4}$ cubic, (Fd$\overline{3}m$) (No. 227), $Z = 8$, 
a = 8.3644(3) \AA \\


\begin{table}
        \begin{tabular}{c|c|c|c|c|c|c|}
        Atom & site & x & y & z  & $B_{iso}$  & Occ.   \\
        Ti1 & 16d & $1/2$ & $1/2$  & $1/2$ & 0.43(8) & 0.75(2)\\
        Li2 & 16d & $1/2$ & $1/2$ & $1/2$ & 0.43(8) & 0.25(2)\\
        Li1 & 8a & $1/8$ &$1/8$ & $1/8$ & 1(1) & 0.9(1)\\
        O1&  32e & 0.2620(5) & 0.2620(5) & 0.2620(5) &0.9(2)& 1  \\
    \end{tabular}
    \caption{LTO blue}
    \label{tab:lto_blue}
\end{table}
Used restrains: Occ(Ti1) + Occ(Li1) = 1 \\ 
Refined Profile Parameters: U=0.14(2), W = 0.0012(3), X=0.17(1), D/L = S/L = 0.239(2) \\

Comment$:$ From compositional refinement the calculation of charge balance reveals a missing charge of $-0.562$ points to additional oxygen vacancies ($\approx$ 7$\%$), not obtainable from XRD refinement. Additionally, the refinement shows Li vacancies ( $\approx$ 10$\%$) on 8a. \\

Phase 2: \ch{TiO2}-rutile (mass fraction $\approx$ $0.90(6)$ $\%$) tetragonal, $P42/mnm$ (No. 136), \\
a = 4.592(4) \AA, c = 2.966(5) \AA \\
Refined Profile Parameters: W = 0.003(10), X=0.4(6), D/L = S/L = 0.239(2) \\

Phase 3: \ch{TiO2}-hollandite (mass fraction $\approx$ $1.18(6)$ $\%$) tetragonal, $I4/m$ (No. 87), \\
a = 10.155(6) \AA, 
c = 2.966(3) Å \\
Refined Profile Parameters: U=0.2(3), Y=0.02(4), D/L = S/L = 0.239(2) \\

\begin{figure}
\includegraphics[scale=1.]{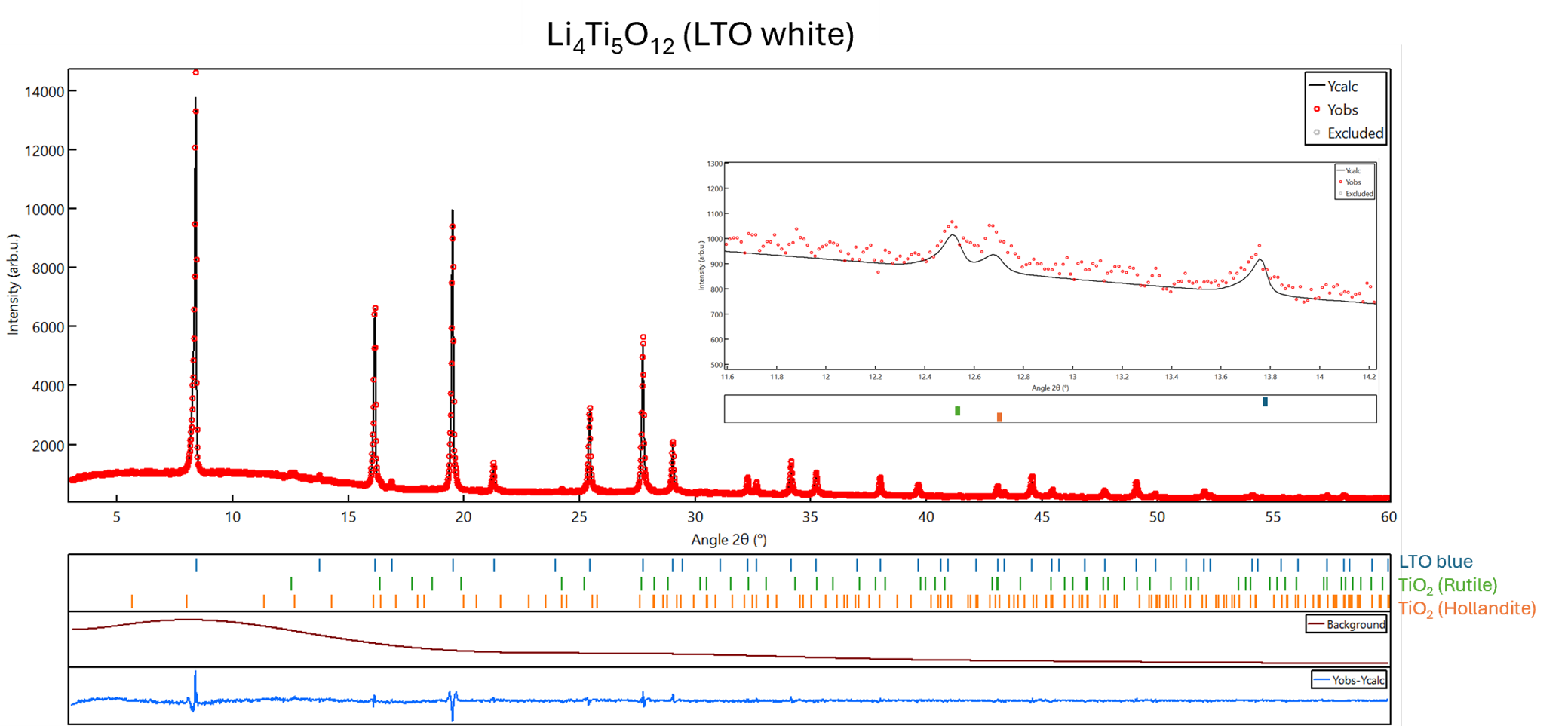} 
\caption{Rietveld refinement of pristine Li$_4$Ti$_5$O$_{12}$ (LTO white) of the X-ray powder diffraction pattern using three phase for refinement. Main phase LTO white. Impurity phases TiO$_2$  rutile-type and TiO$_2$ in an empty hollandite-type.The XRD pattern were recorded with \ch{Mo} $K_{\alpha_1}$ radiation ($\lambda$ = 0.70932 \AA, ~50 $keV$, 40 $mA$)}
\label{WhiteLTO}
\end{figure}

Refinement results: \\
$R_{exp} = 4.22, R_{wp} = 0.0512, Chi^{2} =  1.47 $\\
$R_{Bragg}$ LTO white = 0.0424 \\

Phase 1 $:$ LTO white (mass fraction $\approx$ $98(2)$ $\%$ Composition from refinement $:$ ${Li_{1.526(42)}Ti_{1.474(42)}O_{4}}$
cubic, (Fd$\overline{3}$m),  (No. 227), Z = 8, 
a = 8.3654(2) \AA \\ 

\begin{table}
        \begin{tabular}{|c|c|c|c|c|c|c|}
        Atom & site & x &y	&z& $B_{iso}$	&Occ.   \\
        Ti1 & 16d & $1/2$ & $1/2$  & $1/2$ & 0.52(7) & 0.74(2)\\
        Li2 & 16d & $1/2$ & $1/2$ & $1/2$ & 0.52(7) & 0.26(2)\\
        Li1 & 8a & $1/8$ &$1/8$ & $1/8$ &1.9(10) & 1.01(9)\\
        O1&  32e & 0.2620(5) & 0.2620(5) & 0.2620(5) & 1.0(2)& 1  \\
    \end{tabular}
    \caption{LTO white }
    \label{tab:lto_white}
\end{table}

Used restrains: Occ(Ti1) + Occ(Li1) = 1 \\
Refined Profile Parameters: W = 0.0006(1), X=0.292(8), D/L = S/L = 0.211(2) \\

Comment: From compositional refinement the calculation of charge balance reveals a missing charge of $-0.569$. This points to additional oxygen vacancies ($\approx$ 7 $\%$), not obtainable from XRD refinement. The Li1 occupation refinement indicates no Li vacancies on the 8a site for LTO white. \\

Phase 2 $:$ \ch{TiO2}-rutile (mass fraction $\approx$ $0.8(1)$ $\%$) tetragonal, $P42/mnm$ (No. 136), \\
a = 4.595(3) \AA, 
c = 2.964(3) \AA \\
Refined Profile Parameters: W = 0.003(9), X=0.1(4), D/L = S/L = 0.211(2) \\ 

Phase 3 $:$ \ch{TiO2}-hollandite (mass fraction $\approx$ $1.2(2)$$\%$) tetragonal, $I4/m$ (No. 87), \\
a = 10.140(7) \AA, c = 2.972(4) \AA \\
Refined Profile Parameters: U = 0.3(20), Y=0.05(6), D/L = S/L = 0.211(2) \\

\FloatBarrier


\subsubsection{Positron Annihialtion Lifetime spectroscopy (PALS) and Coincidence Doppler Broadening Spectroscopy (CDBS)}

\begin{figure}[h]
\includegraphics[scale=.25]{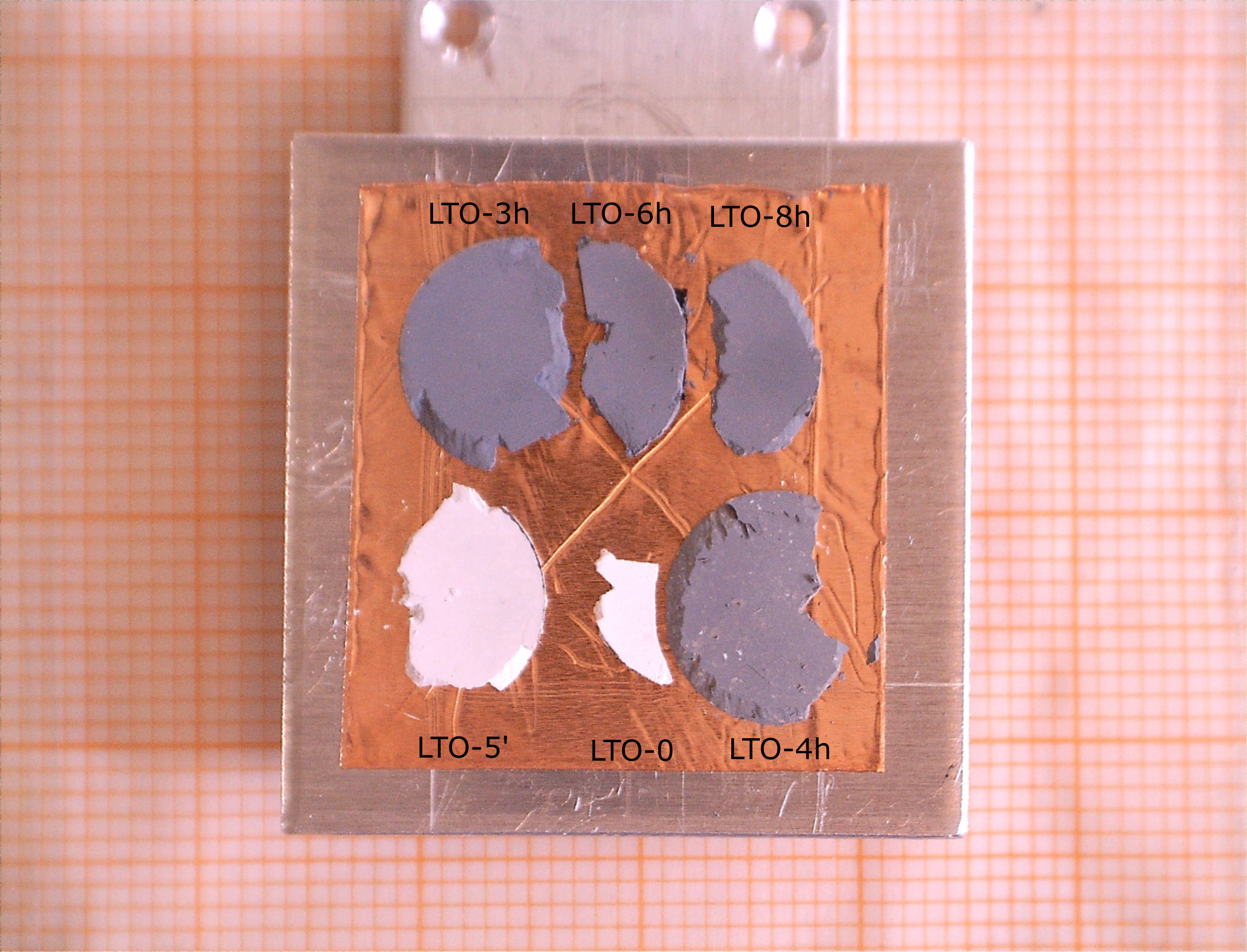} 
\caption{Sample holder for the Coincidence Doppler broadening spectroscopy (CDBS) measurements with the pristine "white" LTO and the "defective" LTO, varying the annealing time. }
\label{Samples_CDBS}
\end{figure}

\begin{figure}[h]
\includegraphics[scale=.15]{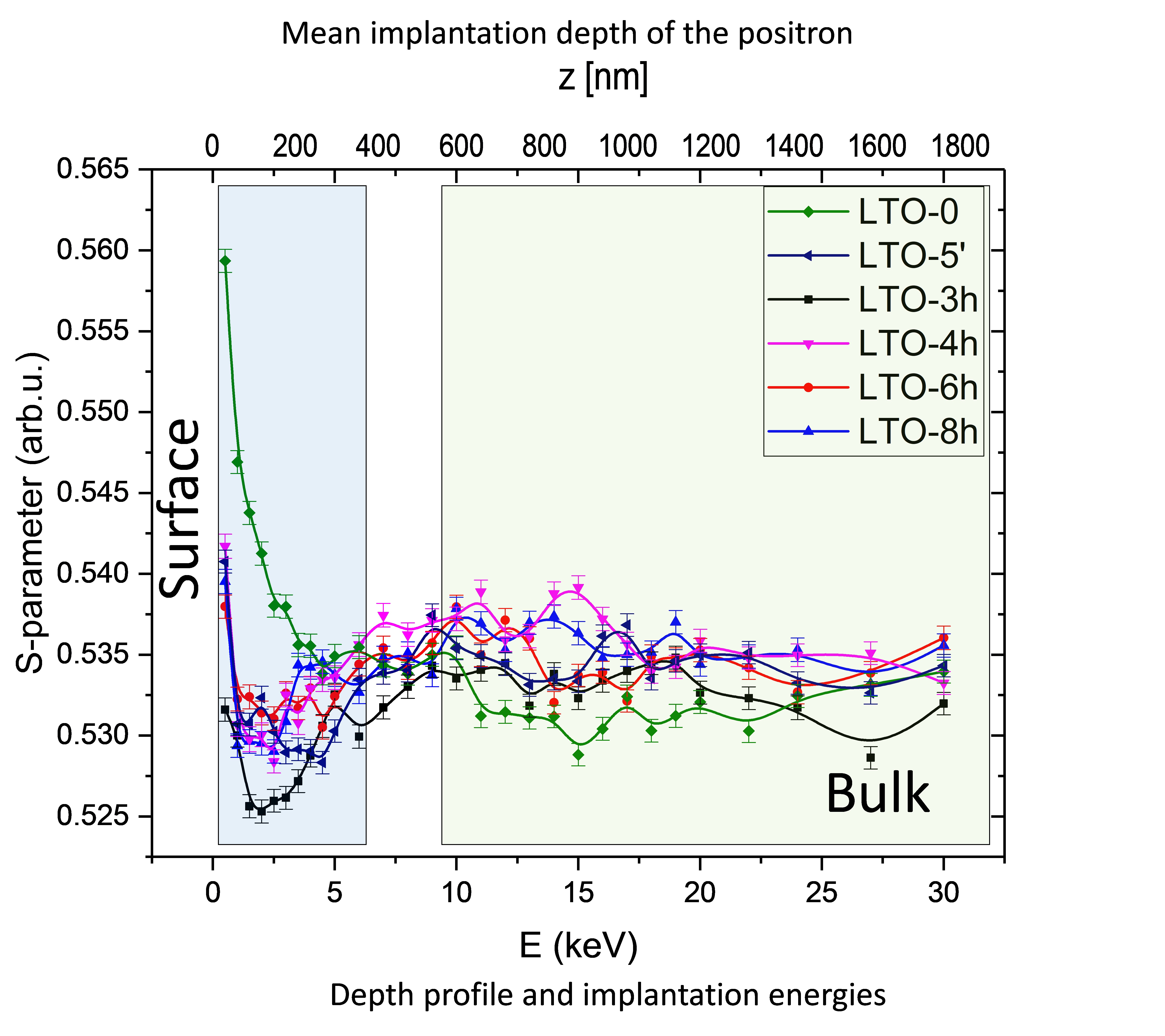} 
\caption{S-parameters scans as function of incident positron energy. Energies within the range of 1-5 keV correspond to surface regions. Energies higher than 18 keV identifies the bulk region/defects.}
\label{SvsE}
\end{figure}

The positron can assess accordingly to the beam intensity a different depth within the material from surface to bulk. The main difference is located within the 0.5-7 keV that goes deep into the surface layer till 5 keV and than continuing within the bulk material 18-20 keV.  
Those it is necessary to scan at first over the beam intensity in order to identify the two energies for surface and bulk measurements for LTO. S parameters is the one scanned over the energies, the fact that decreases till 5 keV Figure \ref{SvsE} indicates a higher concentration of the defect on the surface , confirmed later by the more detailed CDBS.  

\FloatBarrier

\section{Theoretical model}

\subsection{LTO bulk}
\begin{figure}[htbp]
     \centering
     \begin{subfigure}[b]{0.47\textwidth}
         \centering
         \caption{}
         \includegraphics[width=\textwidth]{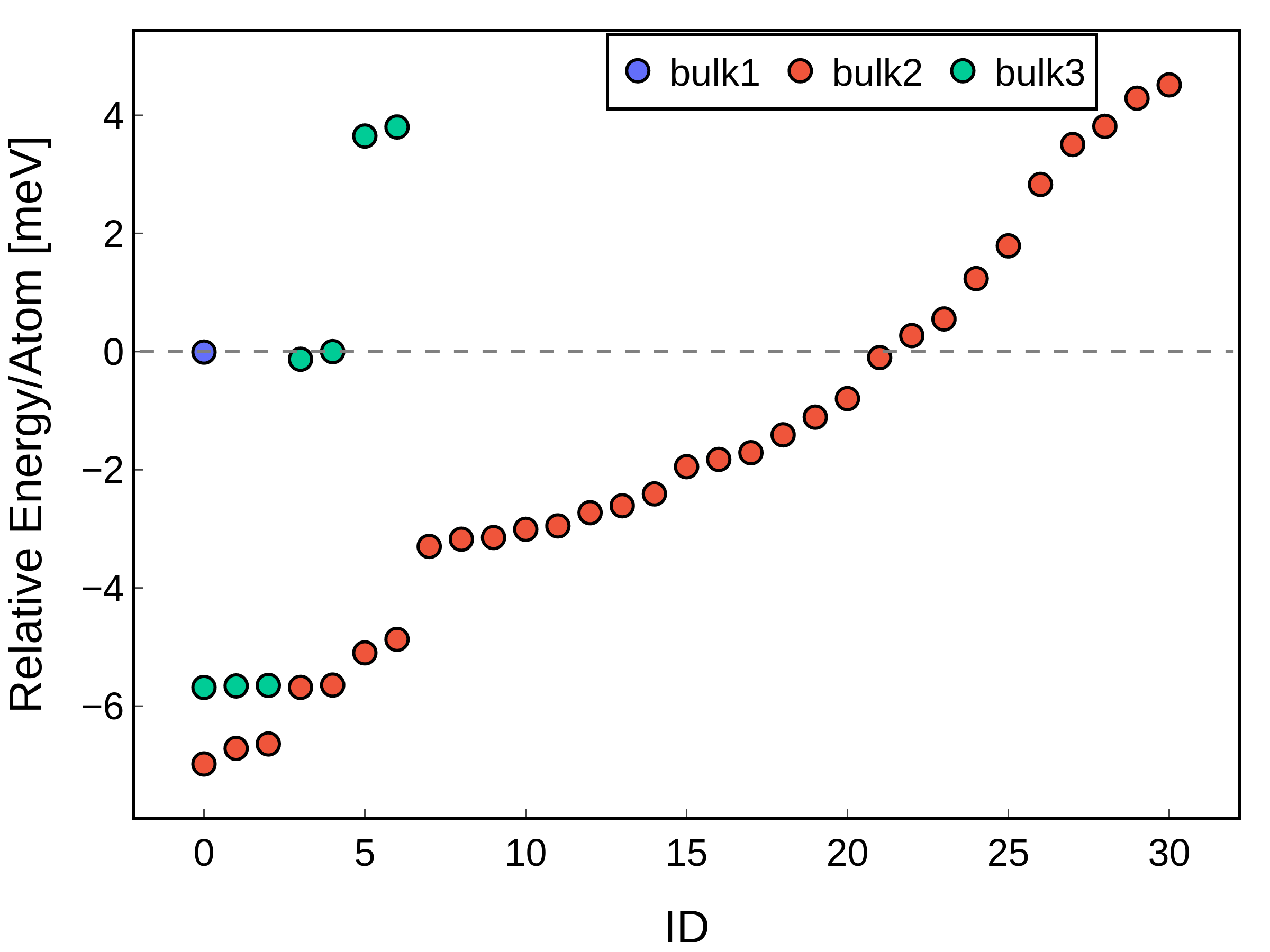}
         \label{fig:bulk formation energy}
     \end{subfigure}
     \hfill
     \begin{subfigure}[b]{0.47\textwidth}
         \centering
         \caption{}
         \includegraphics[width=\textwidth]{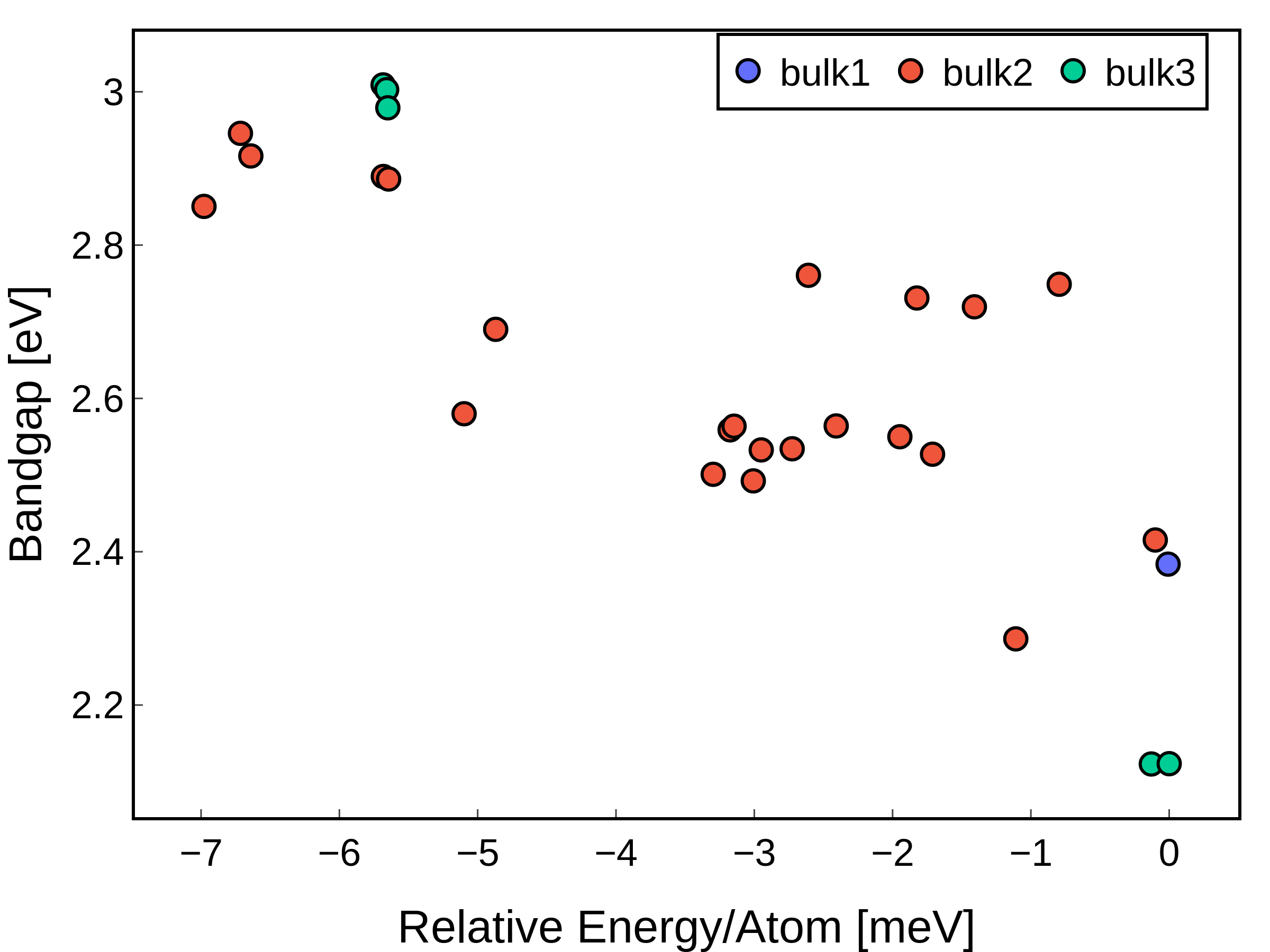}
         \label{fig:bulk bandgap}
     \end{subfigure}
     \hfill
        \caption{(a) Bulk model formation energies and (b) bandgaps}
        \label{fig:Bulk model energies and gaps}
\end{figure}

\begin{figure}
     \centering
     \begin{subfigure}[b]{0.45\textwidth}
         \centering
         \caption{}
         \includegraphics[width=\textwidth]{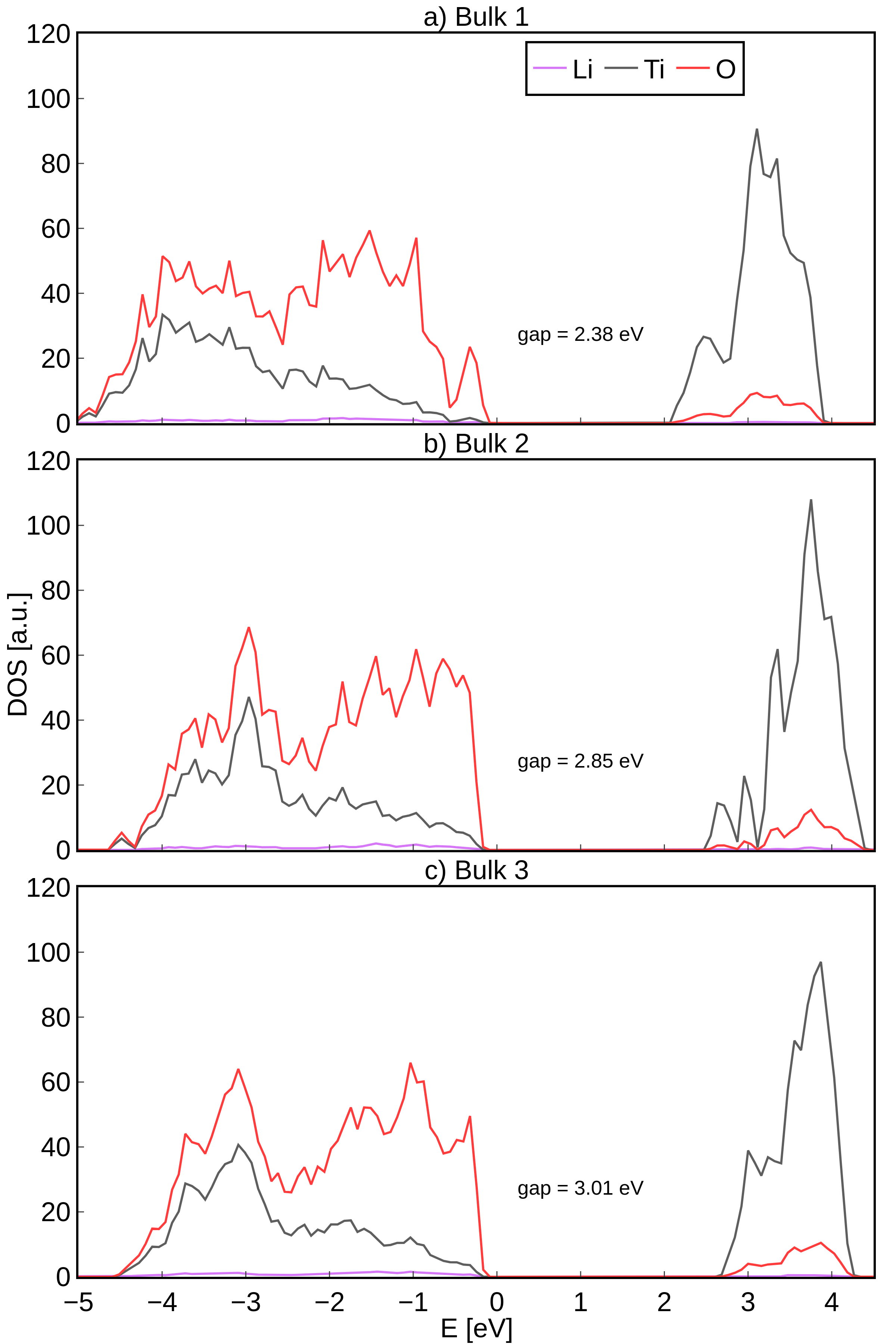}
         \label{fig:PDOS}
     \end{subfigure}
     \hfill
     \begin{subfigure}[b]{0.45\textwidth}
         \centering
         \caption{}
         \includegraphics[width=\textwidth]{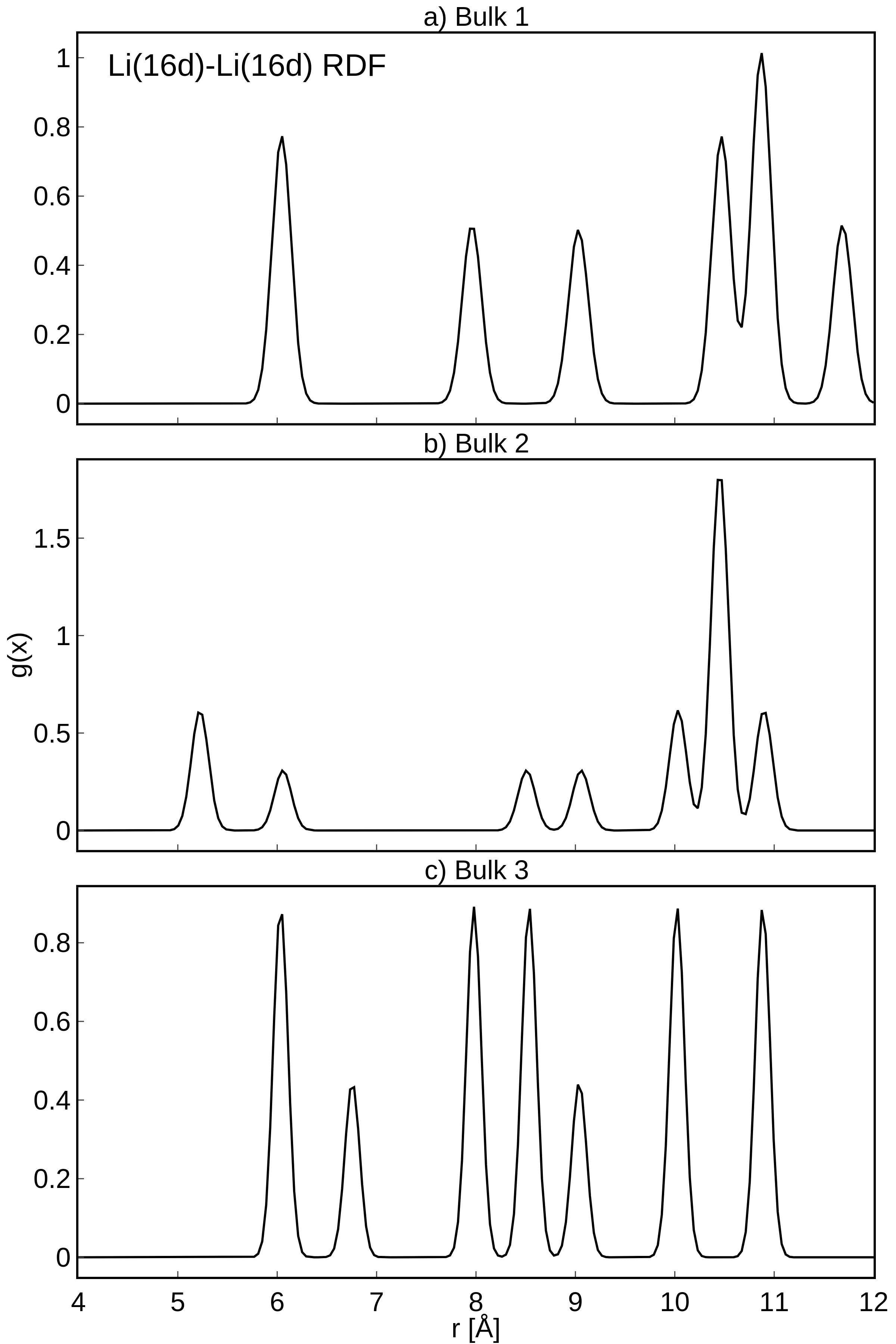}
         \label{fig:RDF}
     \end{subfigure}
     \hfill
        \caption{(a) Bulk model PDOS and (b) radial distribution function}
        \label{fig:Bulk model PDOS and RDF}
\end{figure}
\FloatBarrier

\subsection{Positron-electron interaction benchmark}\label{ssec:Benchmark}
There is no study on LTO positron lifetime both from theoretical and experimental sides. On the other hand, \ch{TiO2} has experimental measurement on positron lifetimes from more than one studies. Zheng et al. studied the positron lifetime change along the \ch{Ruile}-\ch{Anatase} phase transition via temperature control\cite{TiO2_lifetime}. Based on their findings, \ch{Anatase-TiO2} will become \ch{Rutile-TiO2} completely when temperature reach 750 \degree C. The positron lifetime measured on the sample change from 201 $\pm$ 5 ps to 134 $\pm$ 3 ps. Valeeva et al. reported a similar positron lifetime, 148 $\pm$ 4 ps, for \ch{Rutile-TiO2}\cite{TiO2_exp_PL}. Arguably, the same methods which can describe \ch{TiO2} positron lifetimes well should be able to describe LTO since the two out of three elements in LTO are identical (even the oxidation states) with \ch{TiO2}, except \ch{Li}. Moreover, most of the positron density should distribute around oxygen, which attract positron more. In \autoref{tab:ixcpositron_table}, we compare the positron lifetime from different methods on dealing with positron-electron interactions. More details of the electron-positron correlation functional and enhancement factor used in each method can be found on abinit website\cite{abinit_web}. In general, we found that LDA (ixcpositron = 1, 2, 11) gives lifetimes more close to the experimental values for \ch{TiO2}. However, it's hard to describe localized positrons in model with voids, i.e. surface and vacancy models. In fact, based on our experience, LDA can not convergence well for localized systems.  Thus, in the present work, LDA(ixcpositron = 11) is used for systems with delocalized positron (i.e. bulk), and GGA (ixcpositron = 3) is used for localized systems.

\begin{table*}
  \caption{Positron lifetime from different methods.}
  \label{tab:ixcpositron_table}
  \centering
  \begin{tabular}{
    m{\dimexpr 3cm \relax} 
    m{\dimexpr 3cm \relax} 
    *{5}{>{\centering\arraybackslash}m{\dimexpr(\textwidth-14\tabcolsep-5cm)/5}} 
  }
    \hline
 & exp. & & & ixcpositron & &  \\
 &  & 1 & 2 & 11 & 3 & 31 \\
    \hline
\ch{Rutile-TiO2}\cite{TiO2_structures} & 134 $\pm$ 3 ps\cite{TiO2_lifetime}& 154.71 ps & 154.62 ps & 156.50 ps & 177.38 ps & 179.22 ps \\
 & 148 $\pm$ 4 ps\cite{TiO2_exp_PL} & & & & & \\
\ch{Anatase-TiO2}\cite{TiO2_structures} & 201 $\pm$ 5 ps\cite{TiO2_lifetime}& 189.69 ps & 190.28 ps & 191.11 ps & 222.33 ps & 223.86 ps \\
\ch{BBC-Li} & 291 $\pm$ 6 ps& 306.08 ps & 310.49 ps & 303.43 ps & 321.20 ps & 317.86 ps \\
\ch{LTO-bulk}\cite{Li_BCC_structure} &  & 185.86 ps & 186.47 ps & 187.13 ps & 219.70 ps & 220.87 ps \\
    \hline
    \multicolumn{7}{p{\dimexpr\textwidth-2\tabcolsep\relax}}{%
      The k-grids generated by the Monkhorst-Pack scheme used for \ch{Rutile-TiO2}, \ch{Anatase-TiO2}, \ch{BCC-Li} and LTO-bulk are [4, 4, 6], [5, 5, 2], [5, 5, 5], and [4, 4, 2] respectively. More details can be found in Table \autoref{tab:model_info}
    }\\
  \end{tabular}
\end{table*}

\begin{table*}
  \caption{Model information}
  \label{tab:model_info}
  \centering
  \begin{tabular}{
            |c|c|c|c|
  }
    \hline
Model Name & Number of Atoms & Cell Parameters (\AA) & K-Grids \\
\hline
LTO Bulk 1 & 42 & \begin{tabular}{@{}l@{}}(5.973, 0.000, 0.000)\\ (-2.980, 5.161, 0.000)\\ (0.000, 0.000, 14.610)\end{tabular} & 4 x 4 x 2 \\
\hline
LTO Bulk 2 & 84 & \begin{tabular}{@{}l@{}}(6.028, 0.002, 0.006)\\ (0.007, 18.131, -0.021)\\ (0.009, -0.010, 8.535)\end{tabular} & 4 x 2 x 4 \\
\hline
LTO Bulk 3 & 84 & \begin{tabular}{@{}l@{}}(6.035, 0.000, 0.000)\\ (0.000, 6.032, -0.016)\\ (0.000, -0.068, 25.591)\end{tabular} & 4 x 4 x 2 \\
\hline
LTO \ch{V_{Li}} & 167 & \begin{tabular}{@{}l@{}}(12.115, -0.007, 0.000)\\ (-6.051, 10.479, 0.020)\\ (0.000, 0.028, 14.766)\end{tabular} & 2 x 2 x 2 \\
\hline
LTO \ch{V_{O}} & 167 & \begin{tabular}{@{}l@{}}(12.167, -0.007, -0.000)\\ (-6.076, 10.471, 0.014)\\ (-0.000, 0.020, 14.720)\end{tabular} & 2 x 2 x 2 \\
\hline
LTO (111) surface & 168 & \begin{tabular}{@{}l@{}}(11.946, 0.000, 0.000)\\ (-5.960, 10.323, 0.000)\\ (0.000, 0.000, 29.221)\end{tabular} & 2 x 2 x 1 \\
\hline
LTO (110) surface & 168 & \begin{tabular}{@{}l@{}}(8.498, 0.000, 0.000)\\ (0.000, 12.159, 0.000)\\ (0.001, 0.001, 33.536)\end{tabular} & 4 x 2 x 1 \\
\hline
LTO (100) surface & 168 & \begin{tabular}{@{}l@{}}(11.972, 0.007, 0.134)\\ (0.004, 6.039, 0.002)\\ (0.460, 0.015, 41.153)\end{tabular} & 2 x 4 x 1 \\
\hline
\ch{Rutile-TiO2} & 6 & \begin{tabular}{@{}l@{}}(4.594, 0.000, 0.000)\\ (0.000, 4.594, 0.000)\\ (0.000, 0.000, 2.959)\end{tabular} & 4 x 4 x 6 \\
\hline
\ch{Anatase-TiO2} & 12 & \begin{tabular}{@{}l@{}}(3.784, 0.000, 0.000)\\ (0.000, 3.784, 0.000)\\ (0.000, 0.000, 9.514)\end{tabular} & 5 x 5 x 2 \\
\hline
\ch{BCC-Li} & 2 & \begin{tabular}{@{}l@{}}(3.509, 0.000, 0.000)\\ (0.000, 3.509, 0.000)\\ (0.000, 0.000, 3.509)\end{tabular} & 5 x 5 x 5 \\
\hline
  \end{tabular}
\end{table*}
\FloatBarrier

\bibliography{LTO_Positron}